\documentclass[twocolumn,superscriptaddress,citeautoscript,amsmath,amssymb,aps]{revtex4-2}

\usepackage{graphicx}
\usepackage{dcolumn}
\usepackage{bm}
\usepackage{chemformula}
\usepackage{amsmath,amsthm,amssymb}
\usepackage{gensymb}
\usepackage{xr}
\usepackage{hyperref}
\makeatletter
\def\@citess#1#2{\textsuperscript{#1\if@tempswa , #2\fi}}
\makeatother
\usepackage{placeins}
\usepackage{fixltx2e}
\usepackage{upgreek}
\usepackage{soul}  

\usepackage{anyfontsize} 

\usepackage{makecell}

\renewcommand{\figurename}{Figure}  

\makeatletter
\newcommand*{\addFileDependency}[1]{
	\typeout{(#1)}
	\@addtofilelist{#1}
	\IfFileExists{#1}{}{\typeout{No file #1.}}
}
\makeatother

\usepackage[mathlines]{lineno}

\begin{document}
	\setcitestyle{super}
	
	\title{Deterministic nucleation of nanocrystal superlattices on 2D perovskites for light-funneling heterostructures}

	\author{Umberto Filippi}
	\email{umberto.filippi@iit.it}
	\affiliation{Istituto Italiano di Tecnologia, 16163 Genova, Italy}
	\affiliation{University of Notre Dame, Department of Physics and Astronomy, Notre Dame, IN 46556 USA}
	
	\author{Alexander Schleusener}
	\affiliation{Istituto Italiano di Tecnologia, 16163 Genova, Italy}
	
	\author{Simone Lauciello}
	\affiliation{Istituto Italiano di Tecnologia, 16163 Genova, Italy}
	
	\author{Roman Krahne}
	\affiliation{Istituto Italiano di Tecnologia, 16163 Genova, Italy}
	
	\author{Dmitry Baranov}
	\affiliation{Division of Chemical Physics and NanoLund, Department of Chemistry, Lund University, P.O. Box 
		124, Lund, SE-221 00 Sweden}
	
	\author{Liberato Manna}
	\affiliation{Istituto Italiano di Tecnologia, 16163 Genova, Italy}
	
	\author{Masaru Kuno}
	\affiliation{University of Notre Dame, Department of Physics and Astronomy, Notre Dame, IN 46556 USA}
	\affiliation{University of Notre Dame, Department of Chemistry and Biochemistry, Notre Dame,
		IN 46556 USA}
	
	
	\begin{abstract}
		
		Semiconductor heterostructures that combine components with different dimensionality provide an interesting way to manipulate the physical properties of the resulting material. Two-dimensional lead halide perovskites crystallize as flat microcrystals and have efficient in-plane exciton mobility, while perovskite nanocrystals are efficient emitters with a tunable bandgap that can self-assemble into microscopic superlattices. However, combining such intricate architectures into heterostructures has been challenging due to the mismatch in solubility properties and the challenging transfer procedures. Here we realize heterostructures where CsPbBr\textsubscript{3} nanocrystal superlattices are deterministically grown along the faces of PEA\textsubscript{2}PbBr\textsubscript{4} two-dimensional layered perovskite microcrystals. The growth can be limited to the lateral faces of the microcrystals and result in core-crown epitaxial heterostructures, or extended to the vertical direction leading to core-shell-like structures. The growth method is simple yet effective and versatile, and promises to be expanded to a large variety of other materials.
		We demonstrate that these heterostructures can be employed as efficient light-harvesting systems. In fact, energy can be transferred from the two-dimensional microcrystal domain to the superlattices, enabling switching between linear and non-linear carrier recombination regimes by tuning the excitation fluence. Moreover, by exploiting the lifetime shortening of CsPbBr\textsubscript{3} nanocrystal emission upon sample cooling, we ensure that energy transfer occurs after the biexcitonic and single-excitonic decays of the nanocrystals, effectively extending the radiative recombination of superlattices.\\

	\end{abstract}
	
	\maketitle
	\section{\label{sec:level1}Introduction}
	The design of synergistic interactions between different materials is a powerful tool to enhance or transform the properties of individual components that can be accomplished by fabricating heterostructures. Combining different nanomaterials can stabilize one component, boost its performance or even drive the nucleation of a material phase inaccessible by direct synthesis.\cite{li2018photoelectrochemically, zhu2023boosting, toso2022halide} 
	
	Two-dimensional layered metal-halide perovskites (2DLP) are an interesting class of materials to integrate into heterostructures, since the heterojunction interface can be achieved in the in-plane or out-of-plane directions of the layered stacks, which has different effects on energy and charge carrier transport.\cite{schleusener2024heterostructures,pan2021deterministic,shi2020two} 2DLP-based heterostructures made with bulk perovskites of mixed dimensionalities have also been reported, especially in  photovoltaics, where 2DLPs are employed as passivating layers for 3D perovskite crystals to enhance stability.\cite{azmi2025dimensionality,chang2025solvent,la2019vacuum} \textbf{Table \ref{tab:dimension_permutation}} surveys the combinations of mixed-dimensionality perovskite heterostructures reported in the literature.
	
	Heterostructures of mixed dimensionalities incorporating 2DLPs and perovskite nanocrystals have been prepared by epitaxially growing isolated particles atop 2DLP crystals to exploit excitation and charge transfer.\cite{zhu2022room, wu2024real}
	An interesting alternative would be to couple 2DLP microcrystals with nanocrystal superlattices, which are 3D ordered solids made of packed and oriented nanoparticles.\cite{bassani2024nanocrystal}
	Such assembly may result in macroscopic structures with enhanced energy and charge transfer between the donor (2DLP) and the acceptor (superlattice). For instance, multiple arrays of adjacent acceptors would allow to exploit the near-, middle- and far-field coupling regimes, enhancing the overall efficiency of the heterostructure, of which the properties can be finely tuned by engineering the bandgap alignment of the different domains.
	
	The first hurdle to overcome is the heterostructure assembly. Most  protocols for superlattice growth rely on slow solvent evaporation or antisolvent-assisted phase precipitation, which allow nanocrystals to spontaneously aggregate in superstructures.\cite{baranov2019investigation, jana2022self,hiller2025mechanically,baranov2010assembly,levy2024collective} Because of the incompatibility of solubility in a common solvent, nanocrystals cannot be mixed in solution with powders of 2D microcrystals, therefore a post-assembly strategy is required.
	Mechanical manipulation could be used to transfer superlattices in proximity of 2DLP microcrystals and fabricate heterostructures, but this tends to generate irreversible deformation due to the softness of the supercrystals.\cite{nagaoka2017nanocube,hiller2025mechanically} Protocols for the templated growth of particle assemblies have been reported, but they either rely on cumbersome substrate treatments\cite{vila2020templated,kobiyama2025perovskite} or require nanocrystals to merge irreversibly into superstructures.\cite{tong2017precursor}
	\begin{table}[t]
		\centering
		\caption{Perovskite heterostructures with mixed dimensionalities investigated in literature.}
		\label{tab:dimension_permutation}
		\begin{tabular}{c|ccccc}
			& 3D & 2D & 1D & 0D & NCs \\
			\hline
			3D    & $\checkmark$\cite{lin2023all,cui2022liquid}  & $\checkmark$\cite{lu2025layer}  & $\checkmark$\cite{gao20253d}  & $\checkmark$\cite{li2025quantum}  & $\checkmark$\cite{zheng2019quantum}   \\
			2D    &    & $\checkmark$\cite{schleusener2024heterostructures,pan2021deterministic,shi2020two}  & $\times$  & $\times$  & \makecell[l]{%
				$\checkmark$ Isolated NCs\cite{zhu2022room,wu2024real}\\
				$\checkmark$ {Superlattice}$^{(\textrm{This work})}$} \\
			1D    &    &    & $\times$  & $\times$  & \checkmark\cite{rusch2023nanocrystal,patra2024perovskite}   \\
			0D    &    &    &    & $\times$  & \checkmark\cite{chen2018centimeter}   \\
			NCs   &    &    &    &    & \checkmark\cite{livakas2024nanocrystal,cabona2025synthesis,das2025epitaxially}   \\
		\end{tabular}
	\end{table}
	\begin{figure}[t]
		\renewcommand{\figurename}{Scheme}
		\centering
		\includegraphics[width=0.48\textwidth]{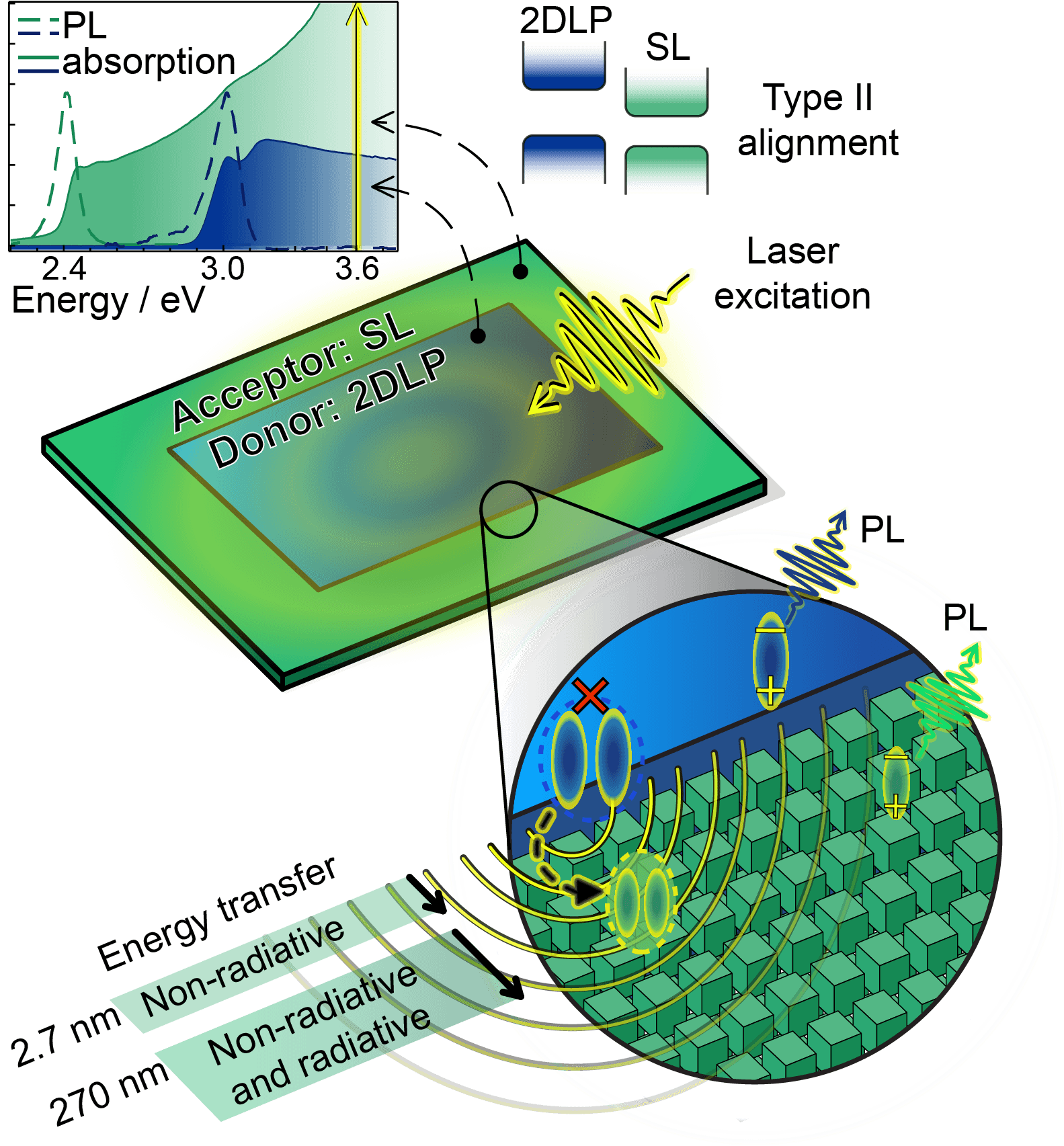}
		\caption{Cartoon summarizing core-crown heterostructures band alignement and optical interaction. After laser excitation, the 2DLP (donor) transfers (non-radiatively and radiatively) the excitation to the superlattice (acceptor).}
		\label{scheme:hts}
	\end{figure}
	\setcounter{figure}{0}
	
	In this work, we report a successful approach to fabricate heterostructures made of CsPbBr$_3$ nanocrystal superlattices and PEA\textsubscript{2}PbBr\textsubscript{4} 2DLP microcrystals. We use 2DLPs as seeds for the heterogeneous nucleation of nanocrystal assemblies during a slow solvent evaporation.  
	Tuning the evaporation time and the nanocrystal concentration results in heterostructures with core--crown and core-shell morphologies.

	We demonstrate the efficient energy transfer from the 2DLP domain to the superlattice, the latter acting as an energy acceptor and radiative component of the system. The efficiency of the energy transfer stems 
	from the exceptionally large F\"{o}rster radius ($\approx$ 67 nm), which enables non-radiative and radiative transfers in the near-, middle-, and  far-field domains, and from the core-shell or core-crown configuration of the heterostructures, where a high bandgap material is surrounded by a superlattice of nanocrystals that feature a smaller bandgap (see \textbf{Scheme \ref{scheme:hts}}).
	
	Moreover, we achieve energy transfer-mediated control over the non-linear phenomena in heterostructures, as the donor domain exhibits a partial suppression of biexciton recombinations due to the depopulation caused by energy transfer.
	As for the acceptor, we identify different temperature regimes to tailor the recombination timescales with respect to the energy transfer, which allows to either accelerate non-radiative biexciton decays or to extend the lifetime of radiative single excitons.\\
	Overall, we demonstrate the capability to assemble powerful light-harvesting systems with switchable optical properties. The remarkable harvesting efficiency is directly linked to the strong light-absorption cross section of the individual building blocks, whereas the transfer efficiency arises from the strong spectral overlap and geometric design of heterostructures (\textbf{Scheme 1}), which are characterized by complete or quasi-complete coverage of the donor domain.
	The combination of these effects transforms the 2DLP microcrystal into an efficient energy funnel that transfers the laser excitation into the superlattices. Further advancements could lead to the development of bio-inspired heterostructures sensitive to extremely low light conditions and capable of funneling the excitation into designated reaction centers.
	
	\section{\label{sec:level1}Results and discussion}
	\subsection*{Heterostructure formation}
	\begin{figure*}[t]
		\includegraphics[width=0.9\textwidth]{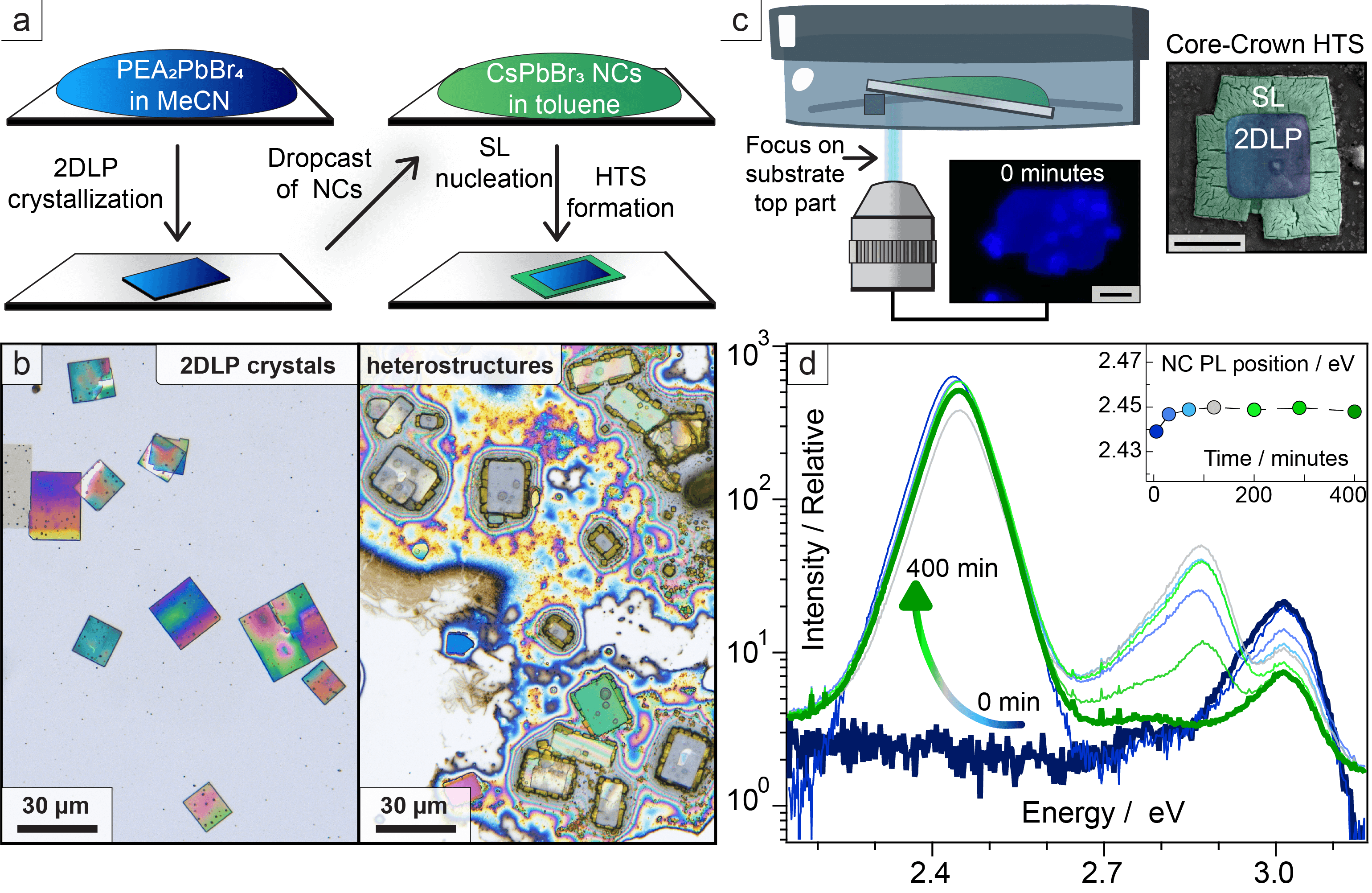}
		\caption{a) Formation of 3D CsPbBr\textsubscript{3} NC SL/2D PEA\textsubscript{2}PbBr\textsubscript{4} crystal heterostructures.  2DLP microcrystals are grown on a designated substrate by an anti-solvent-assisted fast crystallization procedure ($\approx$ 4 hours).\cite{schleusener2024heterostructures} Then a solution of nanocrystals is dropcast onto the same substrate, which is subsequently enclosed in a Petri dish to ensure slow solvent evaporation. After $\approx$ 6 hours, heterostructures are formed.  b) Representative microscope images of regions of the substrate before (left) and after (right) core-crown heterostructure formation. c) Scheme of the inverted microscope set-up used to optically monitor the heterostructures formation (scale bar: 15 $\upmu$m). On the right a colored SEM image of a typical core-crown heterostructure formed on the substrate top part is displayed (scale bar: 15 $\upmu$m). d) \textit{In situ} PL spectra acquired during heterostructure formation. Inset: nanocrystal PL peak position during heterostructure growth.}
		\label{figure:fig_nucleation}
	\end{figure*}
	The preparation of heterostructures is summarized schematically in \textbf{Figure \ref{figure:fig_nucleation}a} and consists of two separate steps: the growth of large $n=1$ PEA\textsubscript{2}PbBr\textsubscript{4} 2DLP microcrystals (see \textbf{Figure S1}),\cite{dhanabalan2019simple,schleusener2024heterostructures,borreani2025direct} and the subsequent growth of superlattices by slow solvent evaporation from a dropcasted nanocrystal dispersion. We used toluene dispersions of CsPbBr\textsubscript{3} nanocrystals passivated with mixed ligands (oleic acid and oleylamine or octylamine, further referred to as C$_{18}$ and C$_{8}$, respectively, see \textbf{Figure \ref{figure:fig_nucleation}b} and Methods).\cite{filippi2025cooling,filippi2026sinusoidal}
	Heterostructure formation occurs on a tilted substrate, which provides a gradient of drying nanocrystal dispersion (the liquid gets progressively concentrated toward the bottom of the film and hence extends the evaporation time in that region, while drying is faster in the upper part, see \textbf{Figure \ref{figure:fig_nucleation}c}). The tilted substrate enables control over the morphology of the resulting heterostructures, as discussed further in the text.
	
	Heterostructure formation was tracked by monitoring the light emission of the constituent materials, starting from individual 2DLP microcrystals located at the top of the substrate (\textbf{Figure \ref{figure:fig_nucleation}d}).
	The intensity of the 3 eV emission peak of the PEA\textsubscript{2}PbBr\textsubscript{4} microcrystal decreases immediately after dropcasting the nanocrystal solution, pointing to optical coupling between the two species.
	
	The PL peak of CsPbBr\textsubscript{3} nanocrystals undergoes a 10 meV blueshift during heterostructure growth, a behavior not observed during growth of pristine superlattices.\cite{baranov2019investigation} According to PL sizing curves, this shift corresponds to a nanocrystal size reduction of $\Delta$NC$_\textrm{size}$ $\approx$ 0.60 nm, which is approximately one CsPbBr\textsubscript{3} unit cell length.\cite{protesescu2015nanocrystals,brennan2017origin} Simultaneously with that change, a new PL peak appears at approximately 2.870 eV, interpreted as a formation of quasi-2D PEA\textsubscript{2}CsPb\textsubscript{2}Br\textsubscript{7} that could be  caused by  residual Cs$^+$ cations transfer to the 2DLP domain.\cite{forlano2023high}  
	The disappearance of the 2.870 eV PL peak at the end of the growth process indicates that this quasi-2D PEA\textsubscript{2}CsPb\textsubscript{2}Br\textsubscript{7} phase does not form a stable interface  neither with the nanocrystals nor the 2DLP domains.
	The growth process  described above is representative for heterostructures growing on the top part of the substrate, where the evaporation time is shorter and the nanocrystal concentration is lower. In contrast, heterostructures forming at the center of the substrate more frequently show a  layer of nanocrystals on top of the 2DLP and retain the PL peak from the mixed phase components even after growth is completed (\textbf{Figure S2}). Complete dissolution of the 2DLP microcrystals can be observed at the bottom of the substrate (\textbf{Figure S3}), where the long evaporation time allows for prolonged dissolution and recrystallization processes.
	The location on the substrate where heterostructures grow also affects the changes in the emission peak intensity of CsPbBr\textsubscript{3} nanocrystals, as it influences the number of emitting nanocrystals and their density (see \textbf{Figure \ref{figure:fig_nucleation}d} and \textbf{Figure S2}).

	\begin{figure*}[t]
		\includegraphics[width=16 cm]{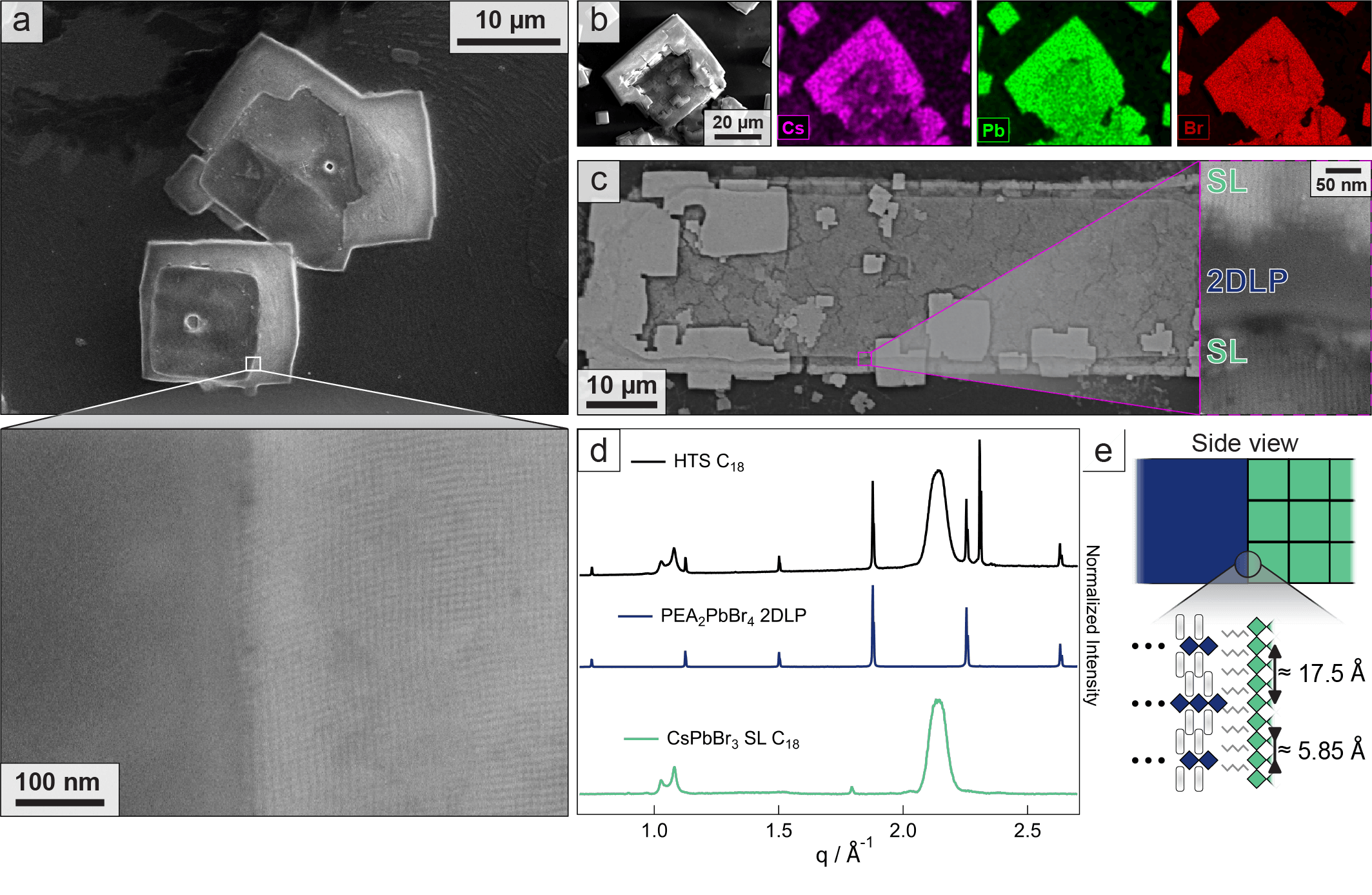}
		\caption{Structural characterization of heterostructures. a) Top: Low-magnification SEM images of the heterostructures. 2DLP microcrystals are located at the center of the structures while superlattices grow at the edges. Bottom: high-resolution image of a representative heterostructure interface, where arrays of nanocrystals are observed.  b) SEM image of one heterostructure and corresponding EDX maps for Br, Cs and Pb. c) Left: Low-magnification SEM image of  heterostructure with fully covered 2DLP microcrystal (nanocrystals assemble also on the top surface). Right: High-resolution image of the heterostructure interface. d)  Comparison between ensemble XRD patterns of pure C\textsubscript{18}-capped CsPbBr\textsubscript{3} superlattices, pristine PEA\textsubscript{2}PbBr\textsubscript{4} microcrystals, and PEA-C\textsubscript{18} heterostructures (black trace on top). e) Cartoon depicting a possible crystallographic alignment between 2DLP and nanocrystals.}
		\label{fig:fig_htsa}
	\end{figure*}
	
	
	\subsection*{Interface characterization}
	
	The \textit{ex situ} morphology of the heterostructures was characterized by scanning electron microscopy (SEM) and x-ray diffraction (XRD) measurements. \textbf{Figure \ref{fig:fig_htsa}a} shows the case of core-crown heterostructures, where CsPbBr\textsubscript{3} nanocrystals assemble with their in-plane facets aligned to the lateral edges of the 2DLP (inset in \textbf{Figure \ref{fig:fig_htsa}a}). Elemental distribution maps confirm that cesium is predominantly localized in the heterostructure regions occupied by superlattices (\textbf{Figure \ref{fig:fig_htsa}b}, see also \textbf{Figure S4} for elemental analysis at the cross section).
	As discussed earlier, the morphology of the heterostructures can be tuned by extending the evaporation time of the nanocrystal solution, allowing nanocrystals to grow also on the top surface of the 2DLP microcrystals (\textbf{Figure \ref{fig:fig_htsa}c}, see also \textbf{Figure S5}).

	Vertical alignment between superlattices and 2DLP microcrystals is likely to occur, as the two components share a similar corner-shared (PbBr$_6$)$^{4-}$ octahedra periodic structure, which leads to a small strain at the interface. Indeed, previous works reported individual nanocrystals exchanging aliphatic ligands with phenethylamine from the 2D component and growing vertically and epitaxially from the top face of the 2D material.\cite{zhu2022room, wu2024real}  
	In contrast,  lateral alignment of the lattices of the two materials is less intuitive due to the mismatch in unit cell periodicities between the two domains. The interlayer spacing  $d$ of the (001) planes of the PEA\textsubscript{2}PbBr\textsubscript{4} microcrystal, calculated from the XRD peak separation as $\Delta q = {1\over d}$, is equal to 16.7 Å (\textbf{Figure \ref{fig:fig_htsa}d}). The first Bragg peak of CsPbBr\textsubscript{3} centered at 1.05 Å$^{-1}$ shows the characteristic splitting of C$_{18}$-capped superlattices,\cite{toso2022collective,toso2021multilayer} and the measured (001) lattice parameter is approximately 5.85 Å (see \textbf{Figure \ref{fig:fig_htsa}e}). Therefore, a possible alignment scenario is that an interface with limited strain can form as three unit cells of CsPbBr\textsubscript{3} nanocrystals ($\approx$ 17.5 Å) match the interlayer distance of the PEA\textsubscript{2}PbBr\textsubscript{4} microcrystal.
	
	An alternative explanation arises from analyzing the heterostructure crystallization process. As discussed in the previous section, the growth involves the partial dissolution of the 2DLP edges (likely driven by the free ligands present in the nanocrystal solution), and Cs ion migration from CsPbBr\textsubscript{3} nanocrystals within the $n=1$ 2DLP edges, leading to the formation of mixed phase domains with $n>1$. These domains generate a mismatch in the periodicity, hence they can easily detach from the main crystal body. These two phenomena  could give rise to ladder-like edges accommodating the nanocrystals both vertically and laterally, which is supported by SEM images taken from the interface of some heterostructures, where alternating regions of nanocrystals and residual 2DLP sheets are observable (\textbf{Figure S6}).\\
	These findings indicate a dynamic reactivity between the 2DLPs and the nanocrystals, involving the exchange of chemical species and resulting in the formation of multiple heterostructure morphologies with different structural properties (see also \textbf{Figures S7, S8 and S9}).
	It is worth mentioning that the 2DLP and the CsPbBr\textsubscript{3} nanocrystals were prepared with different ligand species, and yet they were able to form organized structures. This underpins the versatility of our assembly method and suggests its potential extension to different materials and their combinations.
	In the following of this work, we focus on the investigation of heterostructures by optical spectroscopy to understand the photophysics of both 2DLP and nanocrystal components and their coupling dynamics.
	
	\begin{figure*}[t]
		\includegraphics[width=\textwidth]{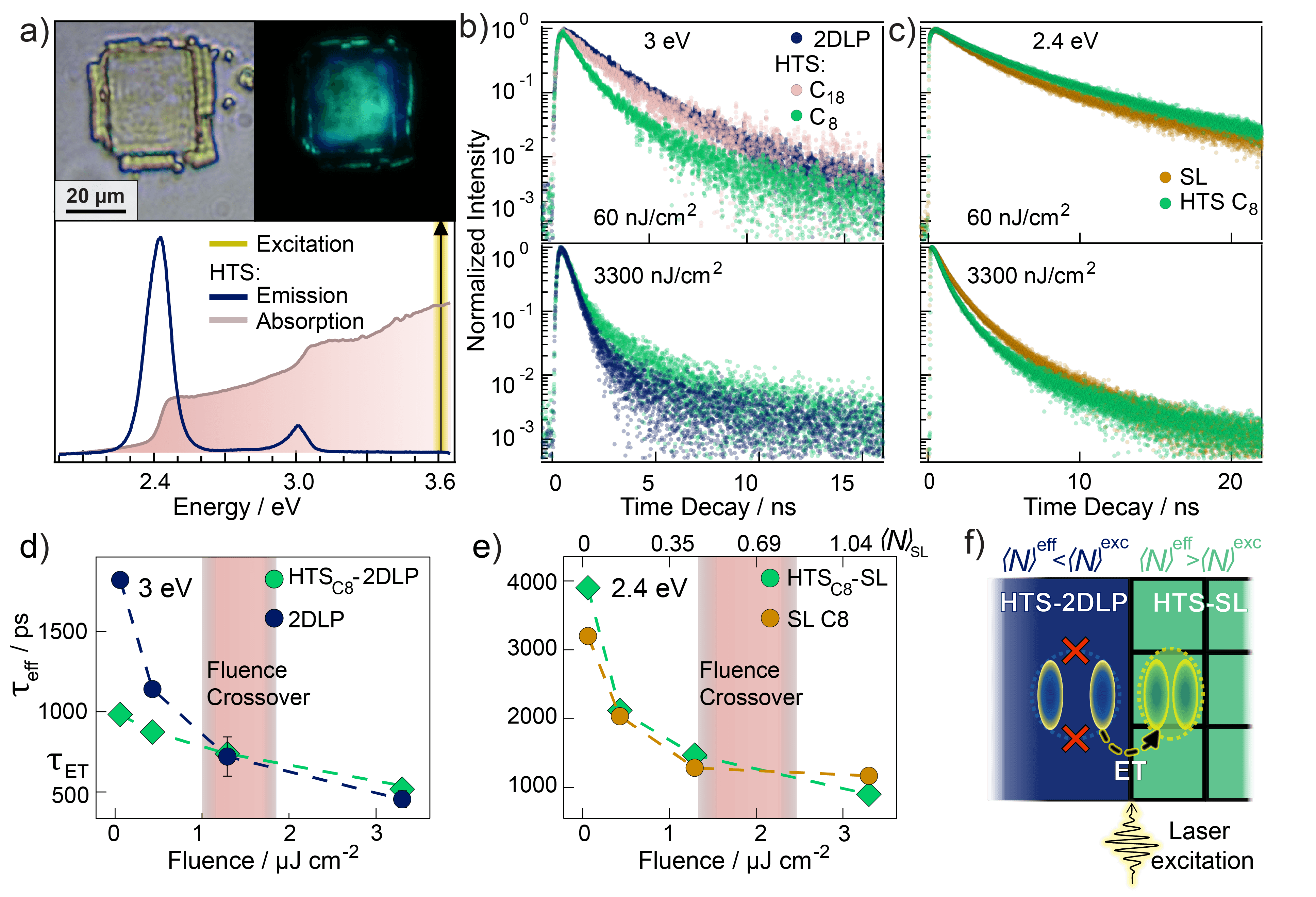}
		\caption{a) Top: Representative microscope images of one heterostructure excited with (left) visible and (right) UV light. Bottom: Absorption spectrum of several heterostructure ensembles and representative PL spectrum of a single heterostructure. b) Time-resolved emission decays of a pristine 2DLP microcrystal and a heterostructure extracted at 3 eV, and collected in the low- (top) and high-fluence (bottom) regimes. c) Similar comparison for traces extracted at the CsPbBr\textsubscript{3} nanocrystal emission peak ($\approx$ 2.4 eV) for a pristine C$_8$ superlattice (orange line) and a C$_8$ heterostructure (spring green line). d, e) Effective lifetimes (weighted average from biexponential fits) as a function of fluence, extracted at 3 eV and 2.4 eV emission energies, respectively. The underlying pink regions indicate the fluence ranges where energy transfer starts to significantly impact biexcitonic recombination relative to single excitonic decays. f) Cartoon depicting the influence of energy transfer on the average exciton population per recombination site  $\langle N \rangle$.}
		\label{figure:fig_LifetimeRT}
	\end{figure*}

	\subsection*{Energy transfer at low fluence}
	
	The literature-reported type II band alignment of CsPbBr\textsubscript{3} and PEA\textsubscript{2}PbBr\textsubscript{4},\cite{zhu2022room, liang2016color, ravi2016band} and the large spectral overlap of the 2DLP microcrystal absorption and nanocrystal PL (see \textbf{Scheme \ref{scheme:hts}}), make these heterostructures a suitable platform for investigating energy and charge transfer processes.
	
	Moreover, the core-shell/core-crown heterostructure geometry combined with the layered architecture of the 2DLP in the center should favor energy transfer: excitons that are created within the 2DLP microcrystal can efficiently diffuse to the edges through the in-plane inorganic layers, where they may be transferred to adjacent superlattices.\cite{seitz2020exciton,seitz2021halide} In addition, the 2DLP emission can be guided laterally  toward the edges of the microcrystal  and reabsorbed by the NC superlattices, as 2DLP microcrystals can act as waveguiding slabs. These processes can be further enhanced by edge states of the microcrystals and their emission.\cite{suarez2025pea2sni4}
	
	Such energy transport and accumulation along macroscopic distances targeting specific recombination centres with nanoscale precision (e.g., the acceptor nanocrystals) could offer a way to potentially control  non linear optical responses (for example by tuning the excitation fluence). However, towards this goal it is fundamental to investigate the timescales on which the different processes occur.
	
	We therefore employed a time-correlated single-photon counting system to measure the time-resolved emission of individual heterostructures upon exciting them with a 343 nm femtosecond pulsed laser. To focus, a 10$\times$ objective lens was employed, resulting in an excitation spot of $\approx$ 9.5 $\upmu$m in diameter. Both 2DLP and superlattice domains were directly excited by the laser beam for the time-resolved emission studies.
	Representative microscope images of one heterostructure and its emission and absorption spectra are shown in \textbf{Figure \ref{figure:fig_LifetimeRT}a}.
	
	To properly address the changes in the optical properties of the heterostructure components, it is essential to characterize them individually. In this regard, we measured the time-resolved emission of pristine 2DLP and superlattices in the low fluence regime, where their recombination is mainly single excitonic (60 nJ cm\textsuperscript{-2}, top graphs \textbf{Figures \ref{figure:fig_LifetimeRT}b,c}).  The lifetimes extracted from single exponential fits are found to be $\tau_\textrm{X,2DLP}=1790$ ps and $\tau_\textrm{X,SL-8}=3200$ ps for pristine 2DLP and superlattices, respectively (the latter assembled with C$_8$-capped nanocrystals). When assembled in heterostructures (HTS$_{{8}}$), the lifetime of the 2DLP domain shortens ($\tau_\textrm{X,HTS$_{{8}}$-2DLP}=980$ ps), while the one of superlattice extends ($\tau_\textrm{X,HTS$_{{8}}$-SL}=3900$ ps). Both phenomena indicate that energy is transferred from the 2DLP to the nanocrystals, as commonly observed in donor-acceptor systems.\cite{rowland2015picosecond,feld2024quantifying,liu2022cu+} 
	
	Within the framework of F\"{o}rster resonance energy transfer (FRET), the rate of transfer is $k_\textrm{ET} \propto (R_0/R)^m$, where $R_0$ is the F\"{o}rster radius, which represents the distance at which energy transfer and donor emission can occur with the same probability, $R$ is the donor-acceptor distance and $m$ depends on the transfer zone (i.e., $m = 6,4,2$ for near, intermediate and far field zone respectively).\cite{clegg2009forster,rogach2009energy,jones2019resonance}
	Because energy transfer includes both non-radiative and radiative contributions that have different scalings, depending on transfer zone, computing $k_\textrm{ET}$ analytically is particularly challenging.  An additional complication arises from accurately estimating $R$, requiring information on the number of donors transferring energy, as well as the number of acceptors distributed across progressively longer distances.

	As energy transfer is expected to occur in a time window smaller than that of 2DLP single exciton decay (i.e., $\tau_\textrm{ET} < \tau_\textrm{X,2DLP}$), we can overcome this issue and estimate an energy transfer lifetime by normalizing 2DLP microcrystal decays in the absence and presence of the acceptor. Normalization is performed at long delay times, where recombination is expected to be predominantly single excitonic.\cite{agarwal2025exciton} This approach results in energy transfer lifetimes of $\tau_\textrm{ET,C8}\approx 640$ ps for C$_8$ heterostructures  (see \textbf{Figure S10}), which increases to $\tau_\textrm{ET,C18}\approx 1000$ ps for C$_{18}$ heterostructures. Therefore, tuning the nanocrystal capping ligands is a tool to control energy transfer timescales.

	\subsection*{Energy transfer at high fluence}
	
	A common strategy to induce non-linear phenomena in nanomaterials is to increase the excitation fluence. 
	This promotes the formation of multi-exciton states, in particular biexcitons, which in confined structures are favored by the strong exciton binding energy enhancing Coulomb interactions. In addition to direct laser photoexcitation, energy transfer can affect the average number of excitons per recombination site $\langle N \rangle$ and influence biexciton formation and recombination. Motivated by the possibility to control non-linear phenomena with energy transfer, we investigated the efficiency of biexciton recombination in heterostructures.
	
	The steady-state PL spectra of pristine 2DLP and superlattices acquired at high excitation fluence (3300 nJ cm\textsuperscript{-2}) do not exhibit any additional emission peak (see \textbf{Figure S11}), which suggests that biexciton recombinations are dominated by non-radiative pathways such as Auger recombination. In agreement with previous reports,  the biexciton lifetime was therefore extracted from time-resolved measurements by performing tail normalization and subtraction between low and high fluence photoluminescence decays of the pristine material (see \textbf{Figures S12,S13}).\cite{agarwal2025exciton,klimov2000optical,rowland2015picosecond,zaffalon2025ultrafast}
	Following this procedure, we measured biexciton lifetimes  $\tau_\textrm{XX,2DLP}\approx 430$ ps and $\tau_{\mathrm{XX,SL}}\approx1000$ ps
	for pristine 2DLP and C$_8$-superlattices, respectively, in good agreement with previous reports.\cite{yin2021auger,makarov2016spectral,zaffalon2025ultrafast} The energy transfer time that we extracted for the heterostructures is $\tau_\textrm{ET,C8}\approx 640$ ps, which is roughly on a similar timescale as the biexciton recombination from the 2DLP domain. As a consequence, energy transfer can partially depopulate the 2DLP biexciton state and effectively decrease biexciton recombination. Indeed, at high fluence, the measured time-resolved emission decay of the donor is overall extended compared to the pristine case (bottom, \textbf{Figure \ref{figure:fig_LifetimeRT}b}), suggesting a reduced contribution from biexciton recombination.
	
	\begin{figure*}[t]
		\includegraphics[width=\textwidth]{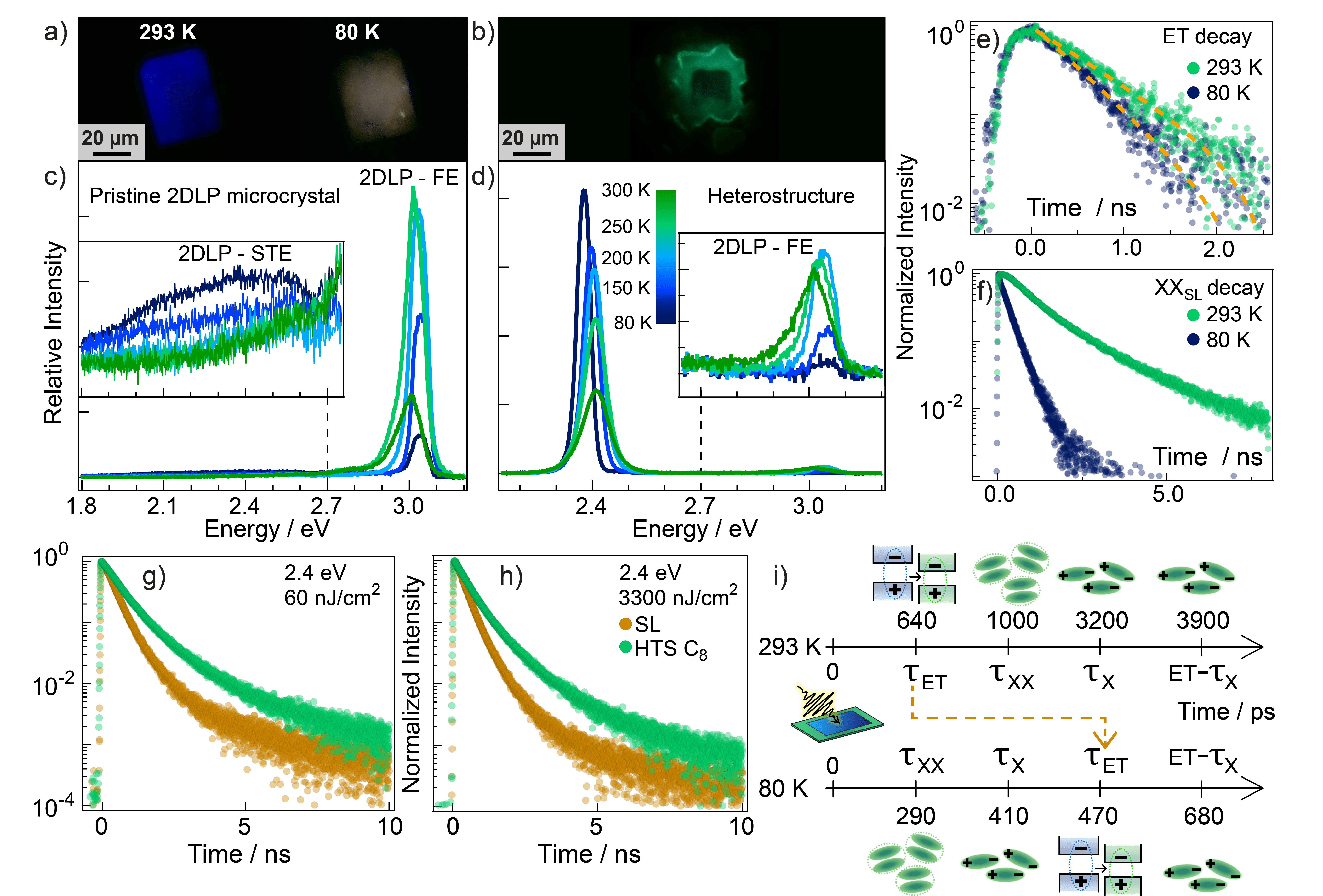}
		\caption{a,b) Representative microscope images of a pristine 2DLP microcrystal at 293 K (left) and 80 K (right) (a) and of a heterostructure (b), both excited with a 343 nm laser. c,d) Temperature-dependent emission spectra of a pristine 2DLP microcrystal and a C\textsubscript{8} heterostructure, respectively. In (c) the inset shows the emergence of the STE at 150 K and in (d) the inset highlights the temperature evolution of the 2DLP peak. e) Energy transfer decays extracted at 80 K and at 293 K from the 2DLP decays (see \textbf{Figure S10}). The orange dashed lines correspond to the best single-exponential fits. f) Biexcitonic decays extracted from the superlattice decays at 293 K and 80 K (see \textbf{Figure S13}). g,h) Time-resolved emission of a heterostructure and of a pristine superlattice extracted at the CsPbBr\textsubscript{3} emission peak in the low- (g) and high-fluence (h) regimes. i) Scheme of all the excitation and recombination mechanisms occurring in the heterostructure in the superlattice domain at 293 K and 80 K.}
		\label{figure:fig_4_lowT}
	\end{figure*}
	
	For the superlattice domain in the heterostructure the situation is reversed, as energy transfer is faster than both single and biexcitonic decays. It hence populates the excited exciton states and increases the effective $\langle N \rangle_\mathrm{HTS-SL}$ (considering only CsPbBr$_3$ nanocrystals and 3300 nJ cm\textsuperscript{-2} as excitation fluence, $\langle N \rangle_\mathrm{SL}$ = 1.16). Therefore the probability to generate biexcitons is higher for superlattices within the heterostructures, and indeed the time-resolved emission decay is shorter compared to pristine superlattices  (\textbf{Figure \ref{figure:fig_LifetimeRT}}c bottom).
	
	To estimate the excitation fluence at which energy transfer starts to affect biexcitonic recombination more strongly than single exciton recombination, in \textbf{Figures \ref{figure:fig_LifetimeRT}d},\textbf{e} we report effective lifetimes extracted from biexponential fits of time-resolved emission decays measured as a function of pump fluence (see \textbf{Figure S14}).
	The effective lifetimes of the donor (measured at 3 eV, \textbf{Figure \ref{figure:fig_LifetimeRT}d}) and the acceptor (measured at 2.4 eV, \textbf{Figure \ref{figure:fig_LifetimeRT}e}) are compared with their pristine counterparts. A crossover is visible for the donor at 1430 $\pm$ 430 nJ cm\textsuperscript{-2} and for the acceptor at 1900 $\pm$ 560 nJ cm\textsuperscript{-2} (pink regions in \textbf{Figures \ref{figure:fig_LifetimeRT}d,e}). For the donor, after the crossover the energy transfer becomes efficient enough to decrease the effective $\langle N \rangle_\mathrm{HTS-2DLP}$ and hence the probability to generate biexcitons. As for the acceptor instead, the effective $\langle N \rangle_\mathrm{HTS-SL}$ increases and the biexciton contribution to the overall decay rises.\\
	This coupling of the decay channels is very interesting, as it suggests that the 2DLP microcrystal acts as an energy funnel that shuffles the laser excitation into the superlattice domain. Such behavior could be exploited for the development of highly efficient light-harvesting nanomaterial systems with unique properties, for example controlled downconversion and FRET-assisted lasing.\cite{klymchenko2025light,okada2018pi,patten2008enhancement} However, under these experimental conditions, the photophysics  of the heterostructures is strongly influenced by non-radiative Auger recombination processes.

	\subsection*{Low-temperature radiative recombination enhancement}
	\begin{table}[h]
		\centering
		\caption{Exciton, biexciton and energy transfer lifetimes at 300 K and 80 K.}
		\label{tab:lifetimes}
		\begin{tabular}{l|cc|cc}
			\hline
			& \multicolumn{2}{c|}{293 K} & \multicolumn{2}{c}{80 K} \\
			\hline
			& $\tau_\mathrm{X}$ [ps] & $\tau_\mathrm{XX}$ [ps] & $\tau_\mathrm{X}$ [ps] & $\tau_\mathrm{XX}$ [ps] \\
			\hline
			2DLP      & 1790 & 430  & 1260 & 350 \\
			HTS--2DLP & 980  & --   & 790  & --  \\
			\cline{1-5}
			SL        & 3200 & 1000 & 410  & 290 \\
			HTS--SL   & 3900 & --   & 680  & --  \\
			\cline{1-2} \cline{2-3} \cline{4-5}
			Energy transfer
			& \multicolumn{2}{c|}{$\tau_{\mathrm{ET\text{-}C_8}} = 640$ ps}
			& \multicolumn{2}{c}{$\tau_{\mathrm{ET\text{-}C_8}} = 470$ ps} \\
			\hline
		\end{tabular}
	\end{table}
	Two strategies may be followed to avoid the dissipation of excitation through non-radiative pathways and exploit energy transfer to induce radiative recombinations at any fluence in the acceptor. First, the acceptor lifetime  can be tuned to be faster than the energy transfer. Alternatively, Auger recombination can be suppressed.\\
	A cooperative method would be to cool the system to cryogenic temperature, which would shorten the lifetime of the acceptor due to the bright ground excitonic state of weakly confined CsPbBr$_3$ nanocrystals,\cite{sercel2019exciton,becker2018bright} as well as suppress phonon populations responsible for Auger efficiency.\cite{pietryga2016spectroscopic,efros2003semiconductor}
	
	\textbf{Figure \ref{figure:fig_4_lowT}a} shows microscope images of a pristine 2DLP microcrystal under 343 nm laser excitation at 293 K (left) and 80 K (right). The 2DLP microcrystal appearance changes from blue to warm white with decreasing temperature.
	The color change originates from the appearance of a broad emission peak centered at 2.4 eV that we assign to self-trapped excitons (STE in \textbf{Figure \ref{figure:fig_4_lowT}b})\cite{milloch2024fate,lin2025carrier,dhanabalan2021engineering}
	that goes along with an intensity decrease of the high energy peak originating  from free exciton recombinations (FE in \textbf{Figure \ref{figure:fig_4_lowT}b}, 3.03 eV at 80 K).\cite{kahmann2021photophysics,van2022temperature}

	In the C$_8$ heterostructure case (\textbf{Figure \ref{figure:fig_4_lowT}c}), the drop of the 2DLP FE peak intensity is even more pronounced (7-fold decrease of the integrated peak area versus 2.5-fold for pristine 2DLP microcrystals, inset in \textbf{Figure \ref{figure:fig_4_lowT}d}). 
	The quenching is rationalized by decreased donor-acceptor separation upon cooling enhancing energy transfer, consistent with the contraction of the interparticle spacing reported for superlattices at 80 K (e.g., a decrease of $\approx$5 Å for C$_8$-capped superlattices).\cite{filippi2025cooling}

	In the temperature range from 293 K to 80 K the measured energy transfer time decreases from $\tau_\textrm{ET}=640$ ps to $\tau_\textrm{ET}=470$ ps (\textbf{Figure \ref{figure:fig_4_lowT}e}, see also \textbf{Figures S12 and S15}), in agreement with our interpretation above. Also, the lifetimes of pristine superlattices shorten compared to room temperature. Specifically, the exciton and biexciton lifetimes at 80 K are $\tau_\textrm{X,SL,80K}=410$ ps and $\tau_\textrm{XX,SL,80K}=290$ ps (see \textbf{Figures \ref{figure:fig_4_lowT}f,g} and \textbf{Table \ref{tab:lifetimes}}).\cite{zaffalon2025ultrafast}
	
	Since at 80 K energy transfer is slower compared to both single and biexciton recombinations in superlattices (i.e., $\tau_\textrm{XX,SL,80K}<\tau_\textrm{X,SL,80K}<\tau_\textrm{ET,80K}$), it cannot contribute significantly to populate the multi-exciton states, i.e. to increase $\langle N \rangle_\mathrm{HTS-SL}$. In agreement with this findings, \textbf{Figure \ref{figure:fig_4_lowT}g} shows that even at high fluence the superlattice emission lifetime is longer in the heterostructure as compared to the individual superlattice, which indicates energy transfer to  single exciton states with long radiative lifetime.   
	
	Also at low temperature no distinct biexciton emission peak is observed up to 3300 nJcm\textsuperscript{-2}, indicating  efficient Auger recombination. However, at low temperature, higher fluence regimes can be explored without compromising the integrity of the heterostructures. At 1.6 mJ cm\textsuperscript{-2}, pristine superlattices display an additional red-shifted emission peak ($\approx$ 30 meV) with superlinear fluence dependence  (\textbf{Figure S16}). This peak, attributed to biexciton emission, is also observable from heterostructures  at approximately the same fluence, and it is characterized by the similar decay dynamics (\textbf{Figure S17}). The similar behavior of the biexciton emission in isolated superlattices and heterostructures is consistent with the timescales reported above: the superlattice biexciton state is not influenced by energy transfer as it occurs on faster time scales. To achieve energy funneling into the biexciton also at low temperatures the energy transfer dynamics need to be accelerated, which could be achieved by engineering the architecture and/or composition of the heterostructures (i.e. towards more efficient FRET processes).

	\section*{Conclusion}

	In this work, we demonstrated the heterogeneous nucleation of CsPbBr\textsubscript{3} nanocrystal superlattices triggered by 2DLP microcrystals. This simple approach leads to 2DLP--superlattice heterostructures of various morphologies. 
	The geometric configuration and spectral overlap provide unique opportunities to exploit energy transfer occurring from the 2DLP domain to the superlattices. The large 2DLP domain collects the light excitation and acts as a funnel that directs the energy to the superlattice domain, enhancing either the biexcitonic or single-excitonic recombination, which can be controlled via the excitation fluence and the sample temperature.
	These heterostructures represent exciting platforms that could lead to the development of efficient light-harvesting systems inspired by natural complexes, in which the photon absorption is performed by arrays of chromophores that redirect the energy to photo reactive centers with nanoscale precision. In addition, it forges the way toward the deterministic design of heterogeneously grown nanocrystal superlattices, as the versatility of the methods suggests that it could be expanded to a large library of 2D materials and nanocrystals.

	\section*{Experimental section}
	
	\textit{Chemical and Reagents}: Lead(II)bromide (PbBr\textsubscript{2}, $\geq$ 98$\%$), cesium carbonate (Cs\textsubscript{2}CO\textsubscript{3}, 99$\%$), hydrobromic acid (HBr, 48$\%$), phenethylamine (PEA, 99$\%$), oleylamine (OLAm, C$_{18}$, 70$\%$), octylamine (C$_8$, 99$\%$), oleic acid (OA, 90$\%$), acetonitrile (anhydrous, 99.8$\%$), ethyl acetate (anhydrous, 99.5$\%$), 1-octadecene (ODE, 90$\%$), toluene (anhydrous, 99.7$\%$) were purchased from Sigma-Aldrich and used without further purification.
	
	\textit{Synthesis of PEA\textsubscript{2}PbBr\textsubscript{4} Microcrystalline Powders}: To obtain 2DLP PEA\textsubscript{2}PbBr\textsubscript{4} microcrystalline powders we followed previously developed protocols with minor modifications.\cite{schleusener2024heterostructures,dhanabalan2019simple} The first step was carried out in a 7 mL glass vial by dissolving 92 mg of PbBr\textsubscript{2} (0.25 mmol) in 200 $\upmu$L of HBr, which was then followed by dilution with 2 mL of acetone. Then, 75 $\upmu$L (0.6 mmol) of PEA (phenethylamine, 99$\%$) were added to the solution to trigger the nucleation of PEA\textsubscript{2}PbBr\textsubscript{4} microcrystals. The mixture was stirred for $\approx 3$ hours to ensure a complete reaction and the microcrystals were then recovered by centrifuging the vial at 6000 rpm for 3 minutes and the supernatant was discarded. The precipitate was redispersed in 2 mL of acetone and centrifuged again under the same conditions. The washing procedure was repeated two more times, and the resulting solid powder was dried under vacuum for 1 hour. The final PEA\textsubscript{2}PbBr\textsubscript{4} powder was stored in a nitrogen glovebox.
	
	\textit{Crystallization of PEA\textsubscript{2}PbBr\textsubscript{4} Microcrystals on Solid Substrates}: A PEA\textsubscript{2}PbBr\textsubscript{4} stock solution was prepared by dissolving 10 mg of PEA\textsubscript{2}PbBr\textsubscript{4} microcrystalline powder in 20 mL of acetonitrile. In a 7 mL glass vial, 2 mL of the stock solution was mixed with 2 mL of toluene, which induces the precipitation of the PEA\textsubscript{2}PbBr\textsubscript{4} microcrystals.\cite{borreani2025direct} The vial was then placed on a metal block pre-heated to 100 $^\circ$C and left for approximately 3 minutes to ensure complete dissolution of the microcrystals. 2 cm $\times$ 2 cm glass or silicon substrates were placed in glass Petri dish and 170 $\upmu$L of solution was dropcast on them. The solution was allowed to slowly evaporate for $\approx 5$ hours.
	
	\textit{Synthesis of CsPbBr\textsubscript{3} Nanocrystals}: The synthesis of C$_8$- and C$_{18}$-capped CsPbBr\textsubscript{3} nanocrystals was performed following previously reported protocols with minor modifications.\cite{filippi2025cooling,filippi2026sinusoidal} In a 20 mL glass vial, 72 mg of PbBr\textsubscript{2} (0.20 mmol) together with 5 mL of ODE, 50 $\upmu$L (for C$_{18}$) or 150 $\upmu$L (for C$_8$) of OA and 1.5 mmol of amine (500 $\upmu$L for C$_{18}$ or 250 $\upmu$L for C$_{8}$). The vial was placed in a metal block on top of a heating plate, preheated to 185 $^\circ$C. The mixture was heated up to 175 $^\circ$C ($\approx 5$ minutes to ensure the dissolution of the PbBr\textsubscript{2}), after which the vial was lifted from the block and fixed with a clamp above it.
	
	The solution was allowed to cool to the desired injection temperature (160 $^\circ$C for C$_{18}$ or 170 $^\circ$C for C$_{8}$), after which 0.5 mL of cesium oleate stock solution was injected. The cesium oleate stock solution was previously prepared in a 40 mL glass vial by dissolving 400 mg of Cs\textsubscript{2}CO\textsubscript{3} (1.2 mmol) in 15 mL of ODE and 1.75 mL of OA (5.5 mmol) at 120 °C under nitrogen for approximately 1 hour. Ten seconds after injection, the reaction was quenched by immersion in an ice-water bath while stirring. The solution was transferred to a plastic tube and centrifuged for 5 minutes at 7000 rpm. 
	
	For C$_{18}$-capped nanocrystals, the supernatant was discarded and the precipitate was centrifuged at 5000 rpm for 4 minutes and the residual liquid was collected with a paper tissue. This cleaning step was repeated once more, after which the precipitate was dissolved in 500 $\upmu$L of anhydrous toluene. For C$_{8}$-capped nanocrystals, the supernatant was collected in a plastic tube and 7 mL of anhydrous ethyl acetate was added to it. The solution was centrifuged for 5 minutes at 7000 rpm and the supernatant discarded. The precipitate was centrifuged two additional times at 5000 rpm for 4 minutes to remove residual liquid, after which the precipitate was dissolved in 500 $\upmu$L of toluene. Both C$_8$- and C$_{18}$-capped CsPbBr\textsubscript{3} nanocrystal solutions were centrifuged at 7000 rpm for 5 minutes to remove any solid aggregate and the supernatant was stored in a glass vial. Before each experiment, the solution was centrifuged to eliminate any solid precipitate.
	
	\textit{Preparation of CsPbBr\textsubscript{3} Nanocrystal Superlattices}: Superlattices were prepared on top of 2 cm $\times$ 2 cm glass or silicon substrates that were previously rinsed with acetone, isopropanol and toluene. Substrates were placed within Petri dishes with a small tilt to ensure both an evaporation rate gradient and a concentration gradient on the substrate. After dropcasting 130 $\upmu$L of the nanocrystal solution, the Petri dish was closed to ensure slow solvent evaporation.
	
	\textit{Preparation of CsPbBr\textsubscript{3} Superlattice - PEA\textsubscript{2}PbBr\textsubscript{4} Microcrystal Heterostructures}: For the assembly of heterostructures, 3 of the same 2 cm $\times$ 2 cm glass or silicon substrates on which PEA\textsubscript{2}PbBr\textsubscript{4} microcrystals were previously grown were placed in large glass Petri dish (diameter of $\approx$ 8 cm). The specific number of substrates was chosen to ensure an appropriate evaporation time. Substrates were slightly tilted by laying one edge on a glass slab. CsPbBr\textsubscript{3} nanocrystal solutions were diluted with anhydrous toluene. For the C$_{18}$ solution, a 1:1 dilution ratio was performed (i.e. same volume of nanocrystal solution mixed with toluene), while for the C$_{8}$ one, the solution-to-toluene dilution ratio was 1:2, as the synthesis typically produced a larger amount of nanocrystals. Then, 130 $\upmu$L of the CsPbBr\textsubscript{3} nanocrystal solution was dropcast on top of each film and the Petri dish was closed. The formation of heterostructures lasted $\approx$ 6 hours, after which substrates were dried under vacuum in a desiccator.
	
	\textit{Diffraction Data Collection}: $\theta$:2$\theta$ XRD patterns were acquired on a Panalytical Empyrean diffractometer equipped with a 1.8 kW Cu K$\alpha$ ceramic X-ray tube and a PIXcel3D 2 $\times$ 2 area detector operating at 45 kV and 40 mA.
	
	\textit{Morphology characterization}: The morphology of the heterostructures was investigated by mean of an optical profilometer ZETA, of a Zeiss GeminiSEM 560 (Zeiss, Oberkochen, Germany) field-emission scanning electron microscope (SEM) equipped with a gun operating at 10 kV acceleration voltage, and a Helios G4 UX Dual Beam SEM instrument operating at 10 kV.
	
	\textit{Equilibrium Optical Measurements}: Micro-PL measurements were performed using an optical imaging and spectroscopy system coupled with an inverted microscope (Nikon, Eclipse Ti). To excite the samples, a femtosecond laser was employed (NKT, Origami). The laser pulses (1030 nm) were first frequency doubled and subsequently tripled by means of beta barium oxide (BBO) second harmonic generation, and third harmonic generation crystals to obtain a 343 nm excitation wavelength. The excitation laser was focused on the sample using an objective (Nikon, 10$\times$/0.25 numerical aperture). The emission was collected with the same objective and was measured using a fiber-based spectrometer (Ocean Insight, USB2000)
	
	\textit{Time-resolved measurements}: Time-correlated single-photon counting measurements were conducted using the frequency-tripled laser focused on the sample by means of a 10$\times$ objective. The emission was collected with the same objective and was directed into a Gemini interferometer coupled to a single photon counting avalanche photodiode (Micro Photon Devices). Signals from the avalanche photodiode were analyzed using a time-correlated single photon counting module (Picoquant, PicoHarp 300).
	
	\section*{Acknowledgements}
	DB acknowledges the support from the European Union (ERC Starting Grant PROMETHEUS, project no. 101039683). Views and opinions expressed are, however, those of the authors only and do not necessarily reflect those of the European Union or the European Research Council Executive Agency. Neither the European Union nor the granting authority can be held responsible for them. MK thanks the National Science Foundation (DMR-1952841) for supporting this work. L.M. acknowledges funding from European Research Council through the ERC Advanced Grant NEHA (grant agreement n. 101095974).
	\section*{Conflict of Interest}
	The authors declare no conflict of interest.
	
	\newpage
	
	\bibliographystyle{apsrev4-2}
	\bibliography{A_main_refs}
	
\end{document}


\author{Umberto Filippi}
	\email{umberto.filippi@iit.it}
	\affiliation{Department of Nanochemistry, Istituto Italiano di Tecnologia, 16163 Genova, Italy}
	\affiliation{University of Notre Dame, Department of Physics and Astronomy, Notre Dame, IN 46556 USA}
	
	\author{Alexander Schleusener}
	\affiliation{Department of Nanochemistry, Istituto Italiano di Tecnologia, 16163 Genova, Italy}
	
	\author{Simone Lauciello}
	\affiliation{Department of Nanochemistry, Istituto Italiano di Tecnologia, 16163 Genova, Italy}
	
	\author{Roman Krahne}
	\affiliation{Department of Nanochemistry, Istituto Italiano di Tecnologia, 16163 Genova, Italy}
	
	\author{Dmitry Baranov}
	\affiliation{Division of Chemical Physics and NanoLund, Department of Chemistry, Lund University, P.O. Box 
		124, Lund, SE-221 00 Sweden}
	
	\author{Liberato Manna}
	\affiliation{Department of Nanochemistry, Istituto Italiano di Tecnologia, 16163 Genova, Italy}
	
	\author{Masaru Kuno}
	\affiliation{University of Notre Dame, Department of Physics and Astronomy, Notre Dame, IN 46556 USA}
	\affiliation{University of Notre Dame, Department of Chemistry and Biochemistry, Notre Dame,
		IN 46556 USA}
	
	\title{Supporting Information for \\ Deterministic nucleation of nanocrystal superlattices on 2D perovskites for light-funneling heterostructures}

	\maketitle

	\section{2DLP microcrystal lateral size as a function of temperature}

\begin{figure*}[h]
	\centering
	\includegraphics[width=0.9\textwidth]{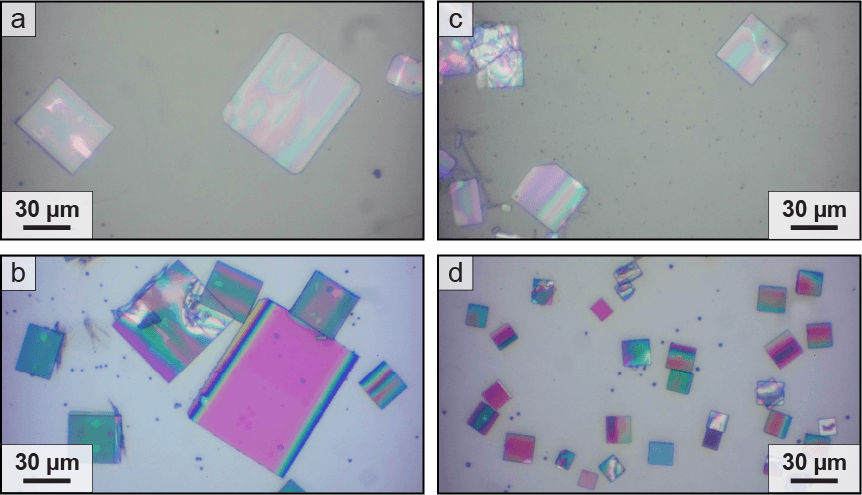}
	\caption{a-d) Microscope images of two-dimensional layered perovskite (2DLP) microcrystals obtained by heating the precursor solution at different temperatures: (a) and (b) at 90 $^\circ$C, (c) and (d) at 60 $^\circ$C. As observable, higher temperature favors the nucleation of larger microcrystals.}
	\label{SI_figure:heating_2dlp}
\end{figure*}

\FloatBarrier
\newpage
\section{\textit{In situ} characterization of heterostructure growth}
\begin{figure*}[h]
	\centering
	\includegraphics[width=0.9\textwidth]{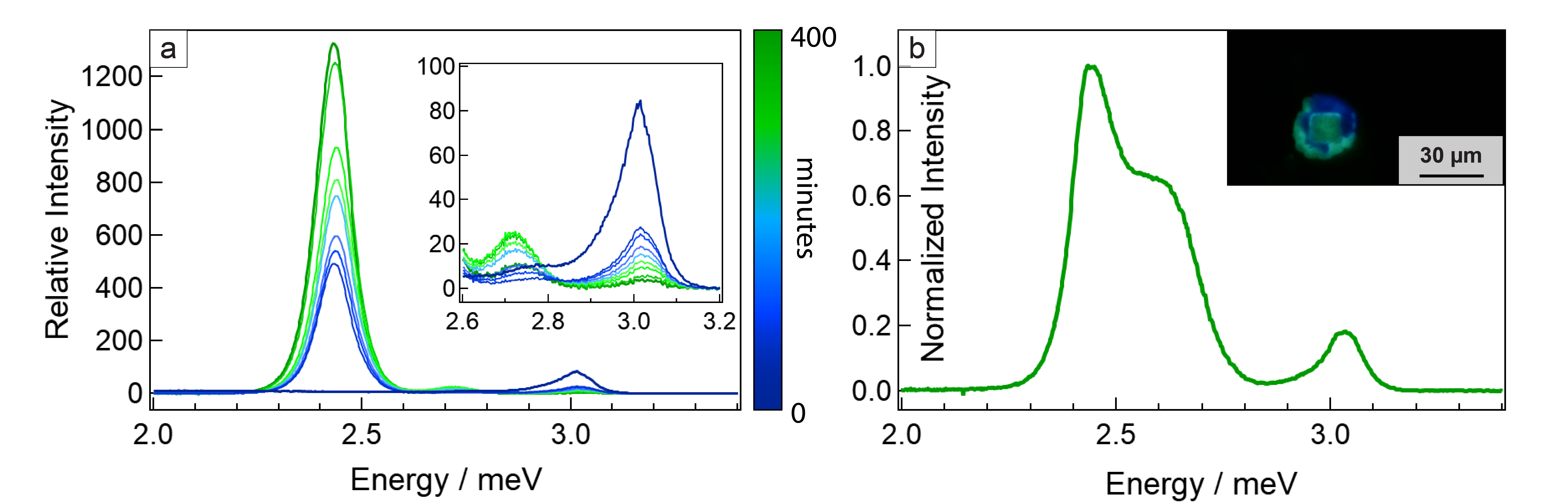}
	\caption{a) Time evolution of the PL spectra during the heterostructure nucleation. Differently from Figure 1c of the main text, the emission at 2.72 eV is still observable after the end of the growth. b) Example of one heterostructure showing the emission peak of a 2DLP microcrystal with n>1.}
	\label{SI_figure:fig_PL_nucle}
\end{figure*}

As specified in the main text, the heterostructure properties depend on the region of the substrate where the heterostructures form. Before dropcasting the CsPbBr$_3$ nanocrystal solution, the substrate, on which the 2DLP microcrystals were previously grown, is tilted in order to create a nanocrystal concentration gradient. As a result, differently from what we reported in Figure 1d in the main text, in the lower regions of the substrate, it is possible to observe heterostructures that retain the emission peak assigned to the quasi-2D microcrystals formed during the heterostructure growth (see \textbf{Figures \ref{SI_figure:fig_PL_nucle}a,b}).\\
Another difference that can be observed depending on the location on the substrate is the change in the emission peak intensity of CsPbBr\textsubscript{3} nanocrystals during the heterostructure formation. The factors affecting this behavior include the number of emitting nanocrystals, the nanocrystal density and the quantum yield.\\
The latter is expected to decrease (independently of the substrate region) as the nanocrystals transition from being dispersed in solution to the solid state (as they precipitate onto the film).\cite{di2017near} In contrast, the number of emitting nanocrystals and their density increase in the lower region of the substrate, as the solution accumulates due to gravity and gradually evaporates. As a consequence, the intensity of the emission from CsPbBr\textsubscript{3} nanocrystals increases with the evaporation time (\textbf{Figure \ref{SI_figure:fig_PL_nucle}a}).  \\
In the upper part of the substrate, the nanocrystal density also increases with time as the solvent evaporates. However, the number of emitting nanocrystals can decrease with time as the solution progressively migrates toward the bottom part of the substrate. Therefore, the decrease of quantum yield and of the number of emitting nanocrystals, make it possible to observe an overall decrease in the emission peak intensity of CsPbBr\textsubscript{3} nanocrystals (as shown in Figure 1d in the main text).

\FloatBarrier

\begin{figure*}[t]
	\centering
	\includegraphics[width=0.9\textwidth]{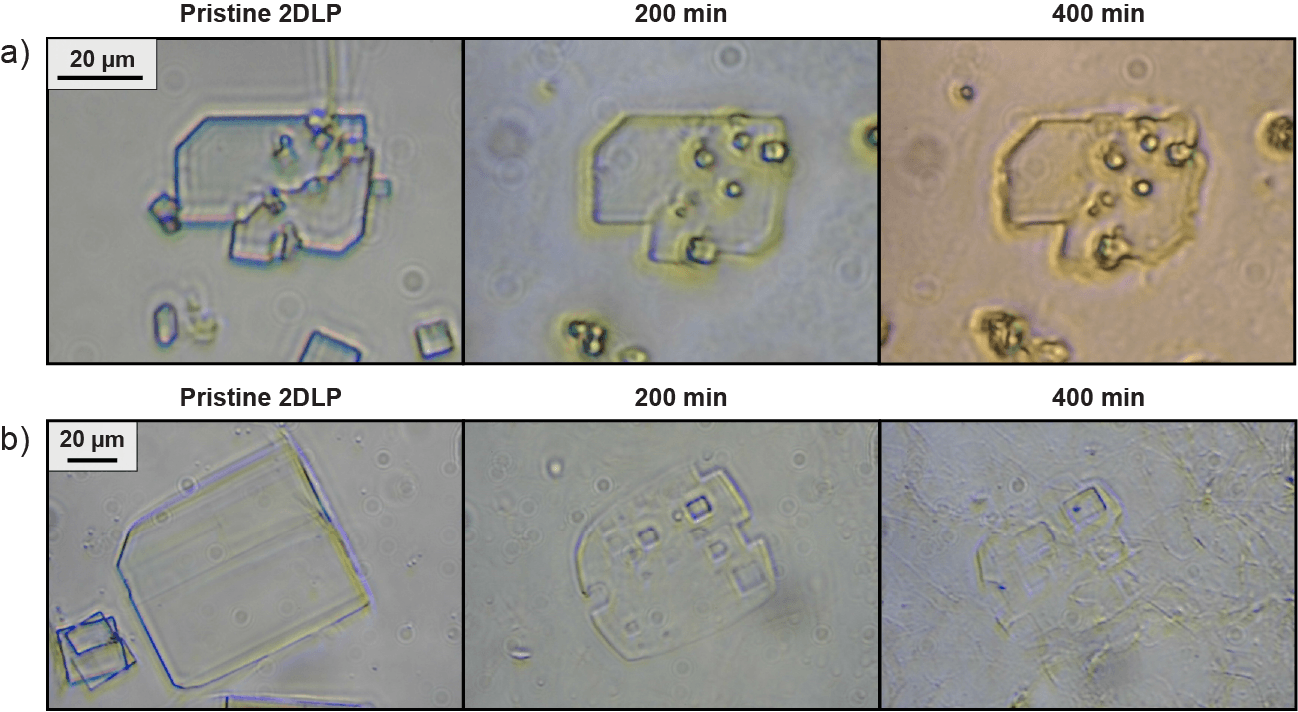}
	\caption{a,b) Time evolution of 2DLP microcrystals after dropcasting the CsPbBr$_3$ nanocrystal solution. In (a). after an initial edge etching, a superlattice domain grows along the edges. In (b), the 2DLP microcrystal is progressively dissolved.}
	\label{SI_figure:fig_dissolution}
\end{figure*}

\FloatBarrier

\section{Heterostructure morphology}
\begin{figure*}[h]
	\centering
	\includegraphics[width=0.95\textwidth]{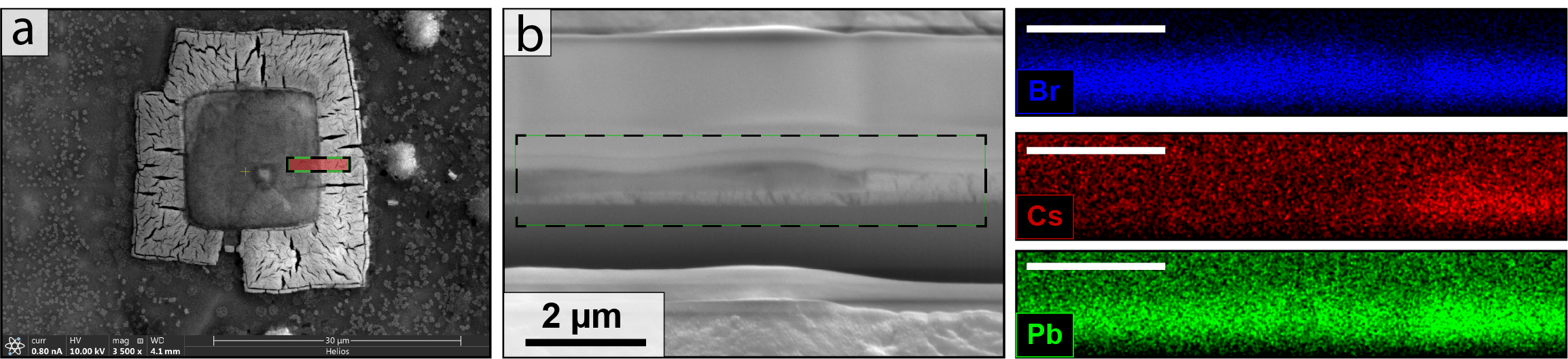}
	\caption{a) SEM image of a core-crown heterostructure. The red box highlights the approximate region removed to obtain the cross-sectional view (box not to scale). b) Cross-sectional image and corresponding EDX maps acquired at the interface of the heterostructures. In the EDX maps the scale bar is 2 $\upmu$m.}
	\label{SI_figure:fig_cross}
\end{figure*}

\FloatBarrier

\begin{figure*}[h]
	\centering
	\includegraphics[width=0.95\textwidth]{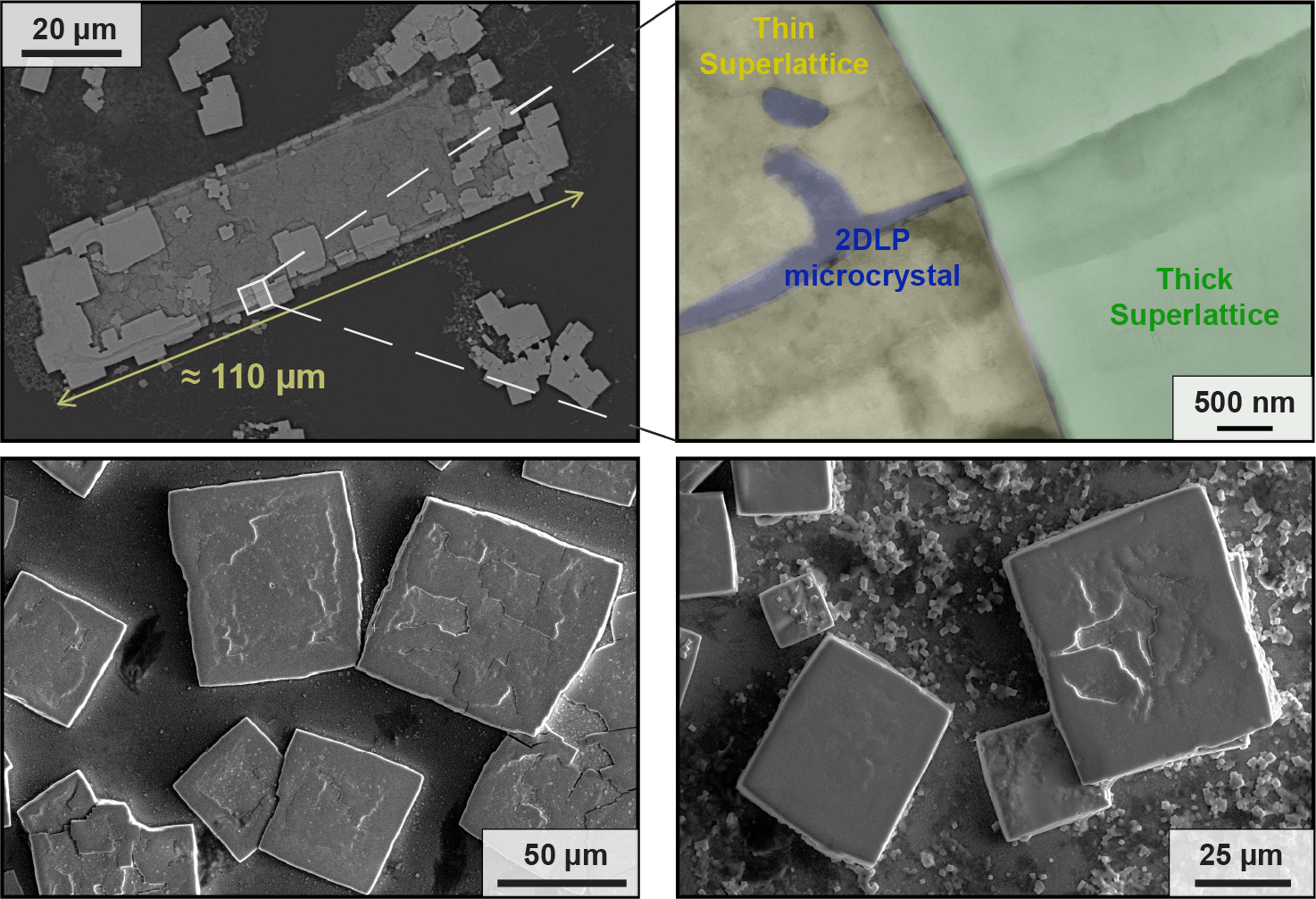}
	\caption{Heterostructure where nanocrystals assembled in correspondence of the lateral and top faces. Large superlattices forming at the edges of the film.}
	\label{SI_figure:large_SL}
\end{figure*}

Under high concentration and long evaporation conditions, that can be achieved for example at the bottom or the edges of the heterostructure film, it is possible to observe heterostructures where nanocrystals assemble also on the top face of the 2DLP microcrystals (\textbf{Figure \ref{SI_figure:large_SL}}, top part). In some extreme cases, the microcrystals fully dissolve, and trigger the formation of large nanocrystal superlattices reaching up to 85 $\upmu$m  (\textbf{Figure \ref{SI_figure:large_SL}}, bottom part). 

\FloatBarrier

\begin{figure*}[h]
	\centering
	\includegraphics[width=0.95\textwidth]{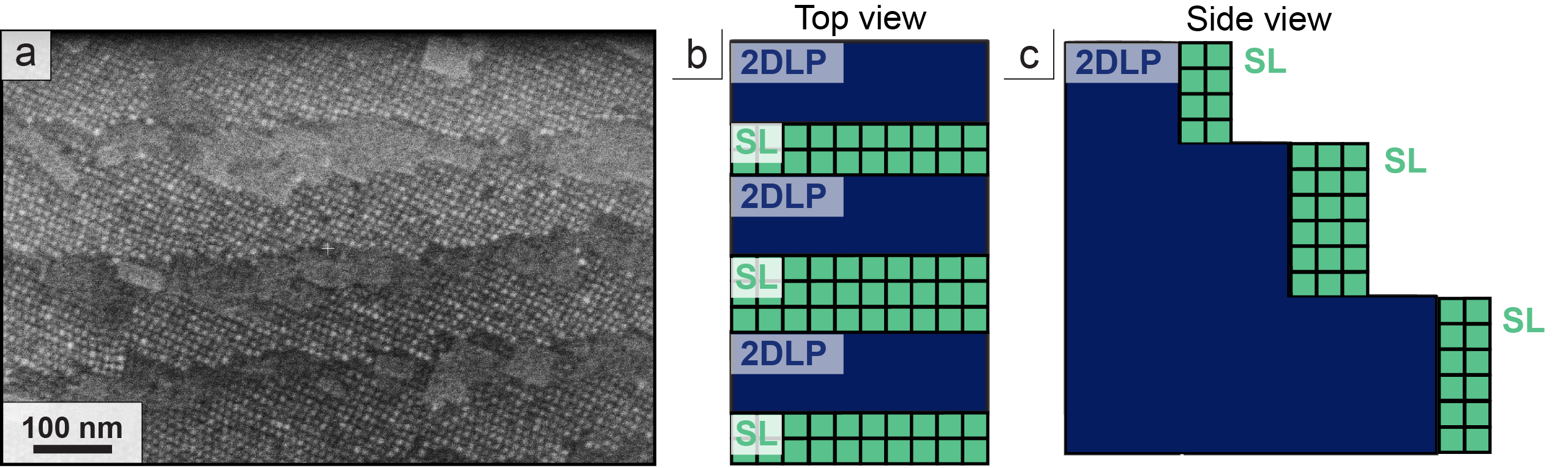}
	\caption{a) Alternating sections of nanocrystals and 2DLP sheets. b,c) Schematic representations of the top (b) and side (c) views.}
	\label{SI_figure:steps}
\end{figure*}

\FloatBarrier
\newpage
\section{Disorder as a function of the heterostructure position on the substrate}

\begin{figure*}[h]
	\includegraphics[width=\textwidth]{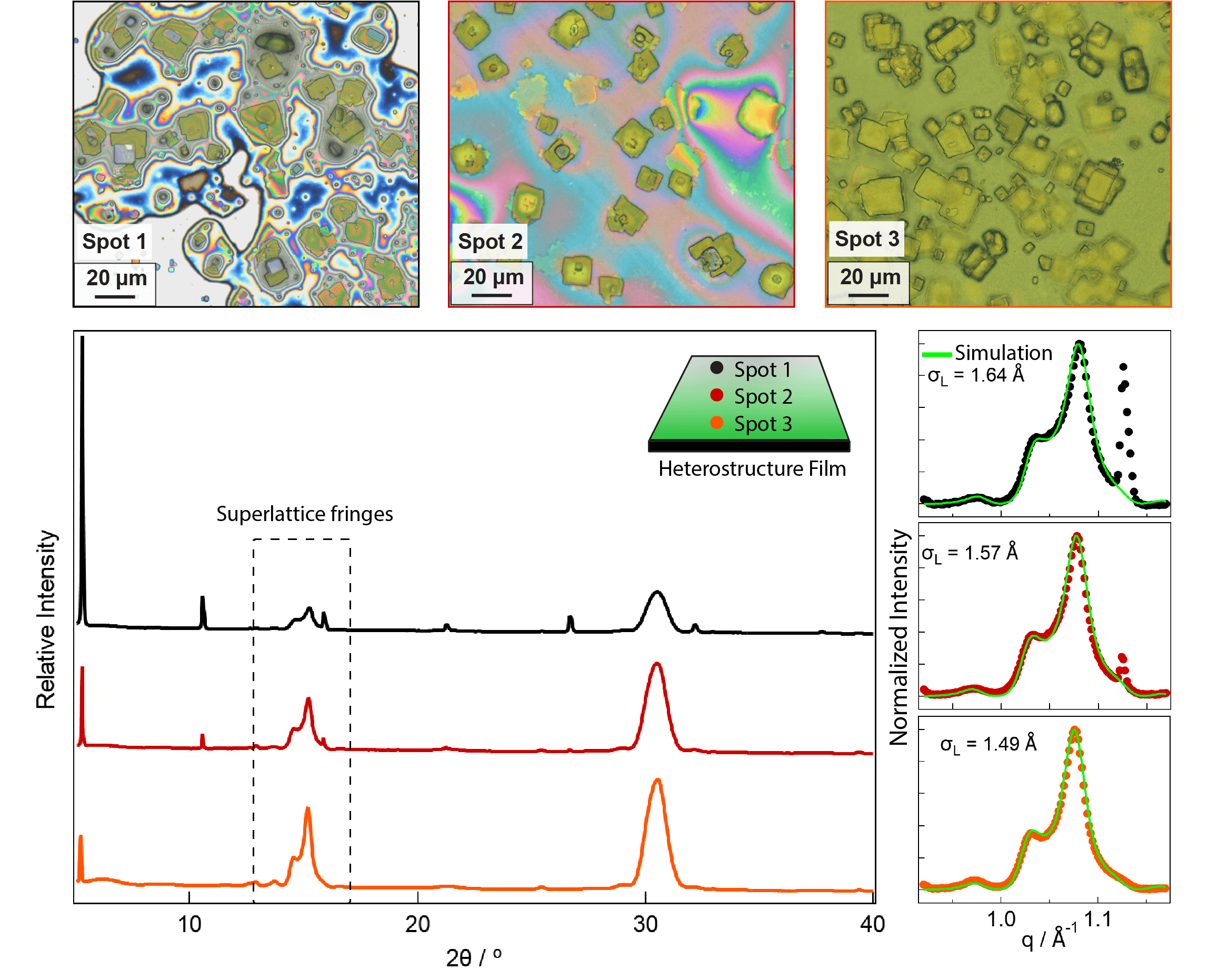}
	\caption{Microscope images and XRD patterns acquired at different positions of the heterostructure film which was prepared by dropcasting the nanocrystal solution on a tilted substrate, to create a concentration gradient. The images show that the top part of the film is characterized by well defined and isolated heterostructures, which progressively become more covered by nanocrystals and eventually the bottom of the film is characterized by a continuous layer of superlattices. The XRD patterns at the same time display broader fringes on the top of the film compared to the bottom, which suggests higher structural disorder.}
	\label{SI_figure:xrd__spots}
\end{figure*}
\FloatBarrier

\begin{figure*}[h]
	\centering
	\includegraphics[width=0.65\textwidth]{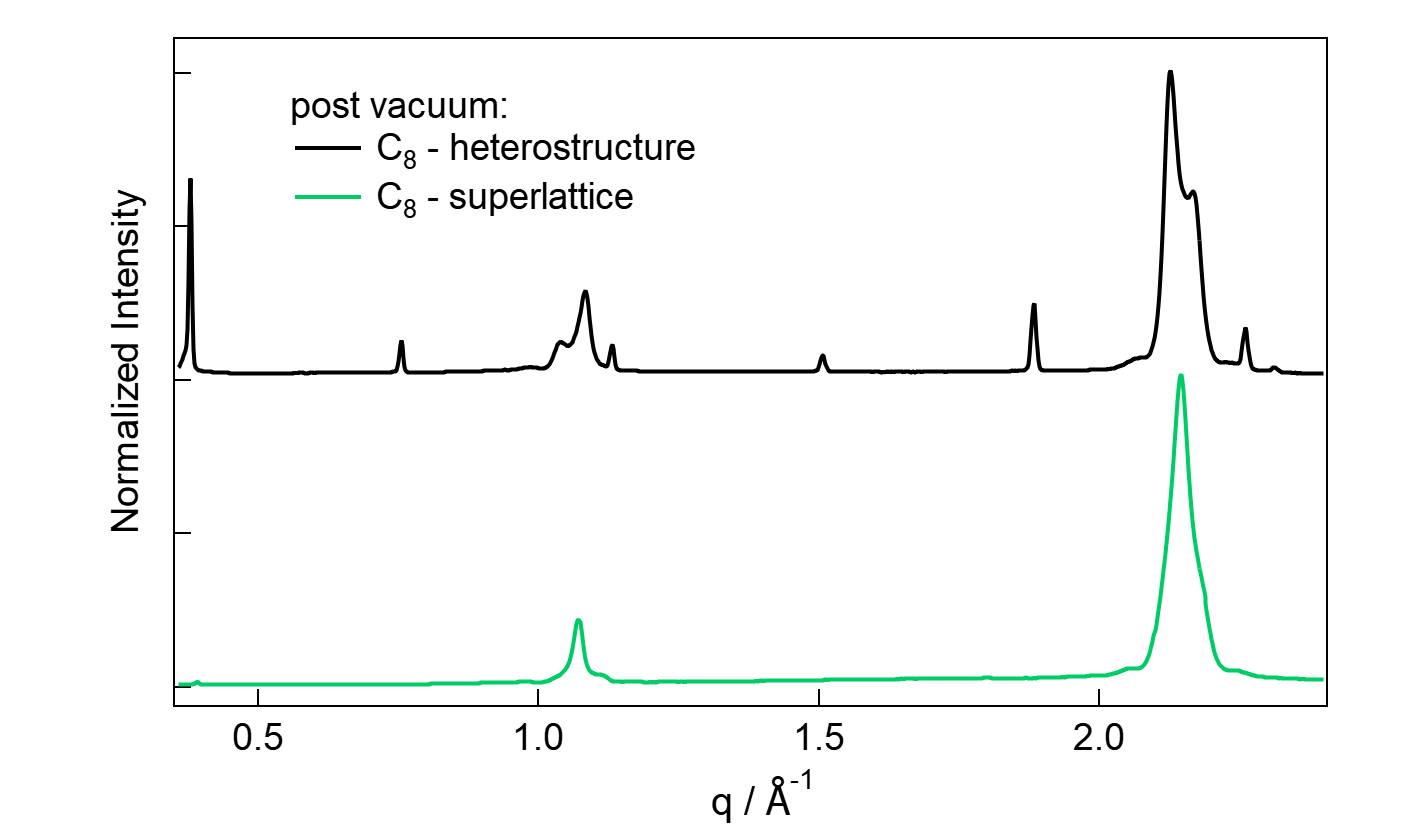}
	\caption{XRD pattern of C$_8$-capped heterostructures and superlattices before and after vacuum. The superlattice fringes in correspondence of the Bragg peaks are lost for the superlattice sample, as some of the volatile ligands are removed when vacuum is applied. For the heterostructure case instead the fringes are still observable, and this suggests that ligand exchange occurs between the 2DLP microcrystals and the superlattices (octylamine exchanged for phenethylamine).}
	\label{SI_figure:fig_XRDC8}
\end{figure*}

To investigate the structural effects induced by these recrystallization and exchange processes on the nanocrystals, we performed XRD measurements along different positions of the heterostructure film.
As we discussed in the main text, because the nanocrystal solution is dropcast onto a tilted substrate, the upper region is characterized by heterostructures formed from a less concentrated solution of nanocrystals, while the lower region experiences larger concentration due to particle accumulation driven by gravity. This difference results in isolated structures in the former region and a continuous film in the latter one.
The XRD signal from the upper domain shows broader superlattice fringes in correspondence of the first Bragg reflection at q = 1.05 \r{A}\textsuperscript{-1} (\textbf{Figure \ref{SI_figure:xrd__spots}}). When fitted using the multilayer diffraction fitting routine developed in previous works for the superlattice analysis,\cite{toso2021multilayer,toso2022collective} it yields a larger value of the average misplacement parameter $\sigma$\textsubscript{L} (1.64 Å vs 1.49 Å). This indicates a greater stacking disorder of the nanocrystals which could arise from the suboptimal superlattice growth condition (as they are force to growth on the 2DLP microcrystals) or from the partial exchange of ligands between 2DLP and nanocrystals (e.g., phenethylamine replacing oleylamine). The latter possibility could represent an intriguing approach to transfer properties between materials. To test these hypotheses, we assembled heterostructures using octylamine-capped (C\textsubscript{8}) CsPbBr$_3$ nanocrystals. This passivation strategy was already employed in our previous work to enhance the structural order of the assemblies, with the drawback of order loss under vacuum, due to the volatility of the ligands.\cite{filippi2025cooling} In the heterostructures case the order is not only enhanced but also it is preserved after vacuum, which can be interpreted as an evidence of partial ligand exchange (C$_8$ with PEA) in the nanocrystals (\textbf{Figure \ref{SI_figure:fig_XRDC8}}).\\

\FloatBarrier

\begin{figure*}[h]
	\centering
	\includegraphics[width=0.4\textwidth]{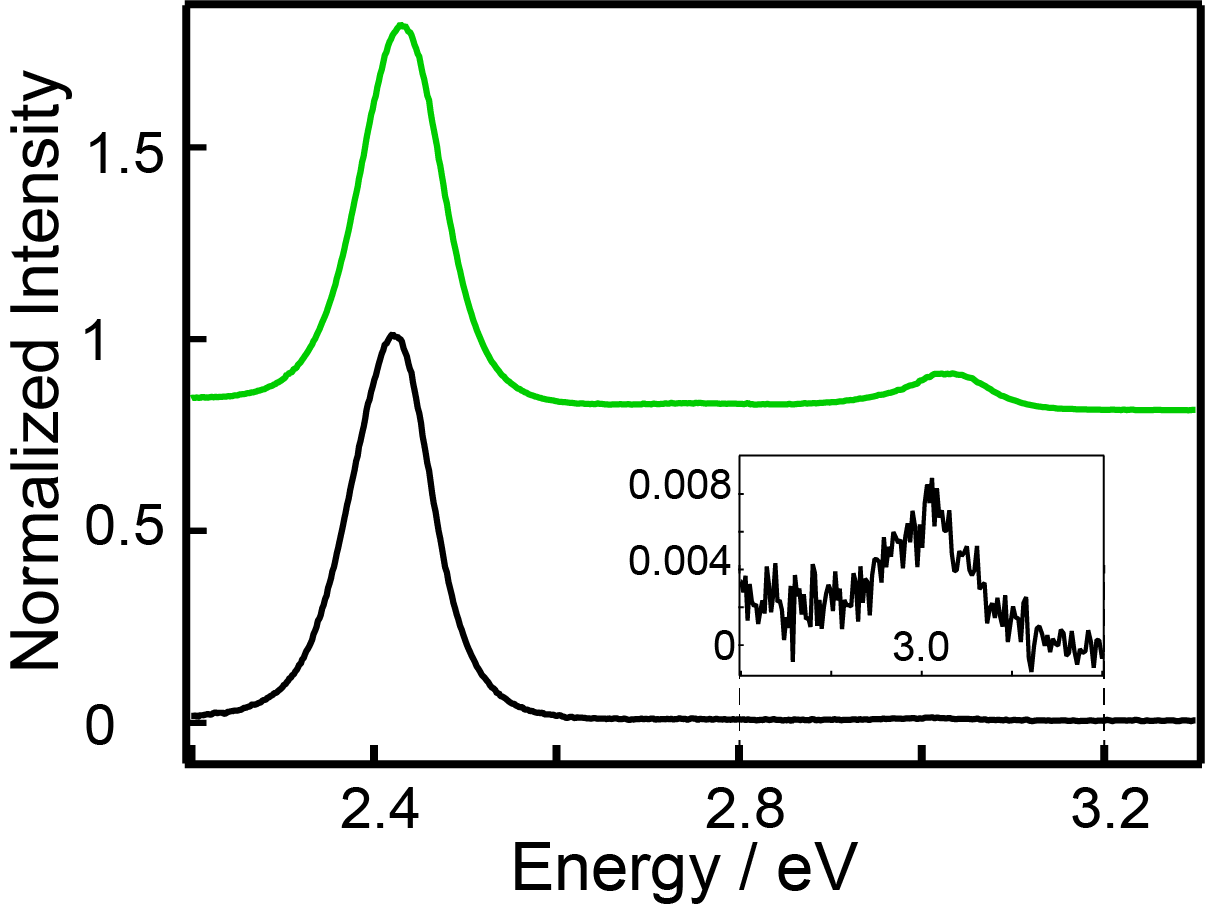}
	\caption{Photoluminescence spectra of two different heterostructures. Top trace shows both the superlattices and 2DLP emission peaks. Bottom spectrum instead exhibit a very weak emission from the 2DLP domain.}
	\label{SI_figure:PL_WEAK}
\end{figure*}

\pagebreak

\section{Calculation of energy transfer parameters}

\begin{figure*}[!htbp]
	\centering
	\includegraphics[width=0.8\textwidth]{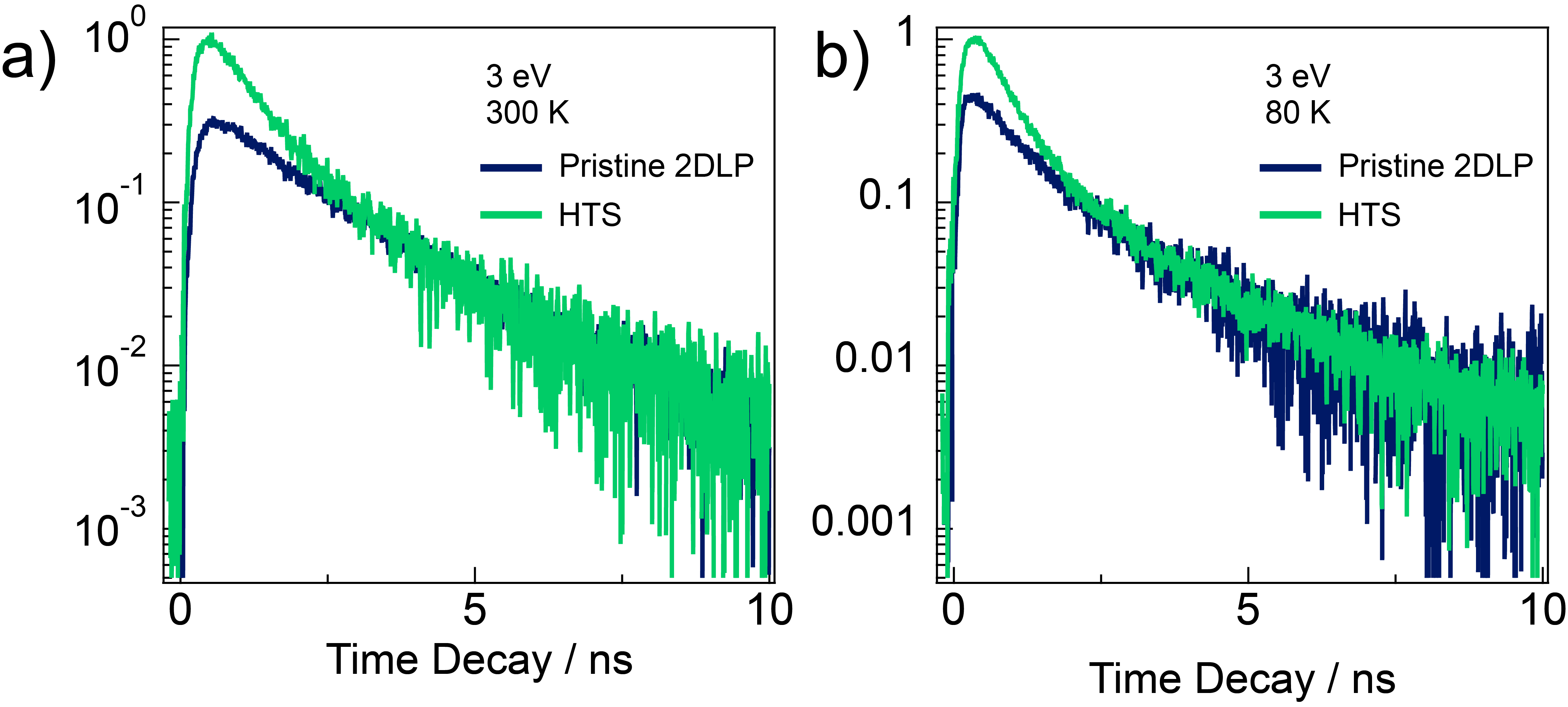}
	\caption{a,b) Time resolved photoluminescence decays measured at 300 K and 80 K respectively for pristine 2DLP microcrystals (dark blue trace) and C$_8$ heterostructures (green trace). The decays are extracted at 3 eV and they are used to calculate the energy transfer decays reported in the main text in Figure 4.}
	\label{SI_figure:fig__ET}
\end{figure*}

The F\"{o}rster radius $R_0$ was calculated using the equation:
\begin{equation}
	R_0 = 0.211 \times\Bigg(\frac{k^2 \eta_{PL}}{n^4} \int_{0}^{\infty}F_D(\lambda)\epsilon_A(\lambda)\lambda^4d\lambda\Bigg),
\end{equation}
where $k$ is the orientation factor, which is assumed to be 2/3 for randomly oriented dipoles, $\eta_{PL} = 0.15$ is the PL quantum yield of the donor in the absence of the acceptor, $n$ is the refractive index of the medium. To estimate $n$, we employed 1.9 for CsPbBr$_3$ (2.2 for the inorganic core and 1.4 for the organic shell, accounting respectively for the 2/3 and  of the 1/3 of final $n$ of CsPbBr$_3$),\cite{feld2024quantifying,zhao2018ellipsometric} and 2.8 for PEA$_2$PbBr$_4$,\cite{chu2023anisotropic} and estimated $n = 1.9\times0.5 + 2.8\times0.5=2.35$. $F_D(\lambda)$ is the PL spectrum of the pristine PEA$_2$PbBr$_4$ 2DLP microcrystal with the integrated area normalized to 1 and $\epsilon_A(\lambda)$ is the absorption of the acceptor expressed in $M^{-1} cm^{-1}$.\\
The integral represents the spectrum overlap integral denoted as $J(\lambda)$ and it results to be $J(\lambda)=$2.457$\times 10^{17}$ $M^{-1}\,nm^4\,mol^{-1}$. The value is particularly large compared with values calculated for example for CsPbBr$_3$ nanocrystals and organic dyes or ligands (5-6 nm)\cite{hardy2021energy,feld2024quantifying,yamaguchi2025energy}. This can be explained by the complete spectral overlap of the PEA$_2$PbBr$_4$ 2DLP microcrystal emission with the CsPbBr$_3$ absorption. The resulting value for the F\"{o}rster radius is $R_0 \approx$67 nm.\\
As discussed in the main text, a large $R_0$ suggests that energy transfer occurs from the 2DLP microcrystal to many nanocrystals vertical arrays. Therefore, energy transfer will include different zones, and its nature will change depending on which zone we are considering. The zones are defined with respect to a distance $b\sim\frac{\lambda}{2\pi n}$, where $\lambda$ represents the donor emission wavelength.\cite{van2013forster} The non-radiative energy transfer should occur in the near-field zone between $0.01b$ and $0.1b$ ($\approx$ 2.7 nm) and scale with $m = $6. The intermediate-zone should extend until $10b$ ($\approx$ 270 nm), scale with $m=$4 and include both non-radiative and radiative contributions. The far-zone is the area extending beyond $10b$ $\approx$ 270 nm, and it is dominated by radiative energy transfer.\\
We can calculate the energy transfer rate $k_{ET}$ as, 
\begin{equation}
	k_{ET} = \frac{1}{\tau_{ET}},
\end{equation}
where $\tau_{ET}$ is the energy transfer lifetime, which we estimated by normalizing the time-resolved emission of the pristine 2DLP microcrystal at long decay times, where recombination is expected to be predominantly single excitonic, and eventually subtracting from it the 2DLP microcrystal decay when assembled in the heterostructures (previously normalized with the same approach).\\
The normalized traces are shown in \textbf{Figure \ref{SI_figure:fig__ET}} and the calculation leads to an energy transfer lifetime $\tau_{ET} \approx$ 640 ps and 1000 ps for the C$_8$ and C$_{18}$ heterostructures respectively. We can employ these values to estimate the energy transfer efficiency.

\FloatBarrier
\newpage
\section{Fluence Dependent Photoluminescence Spectra}
\FloatBarrier

\begin{figure*}[h]
	\centering
	\includegraphics[width=0.85\textwidth]{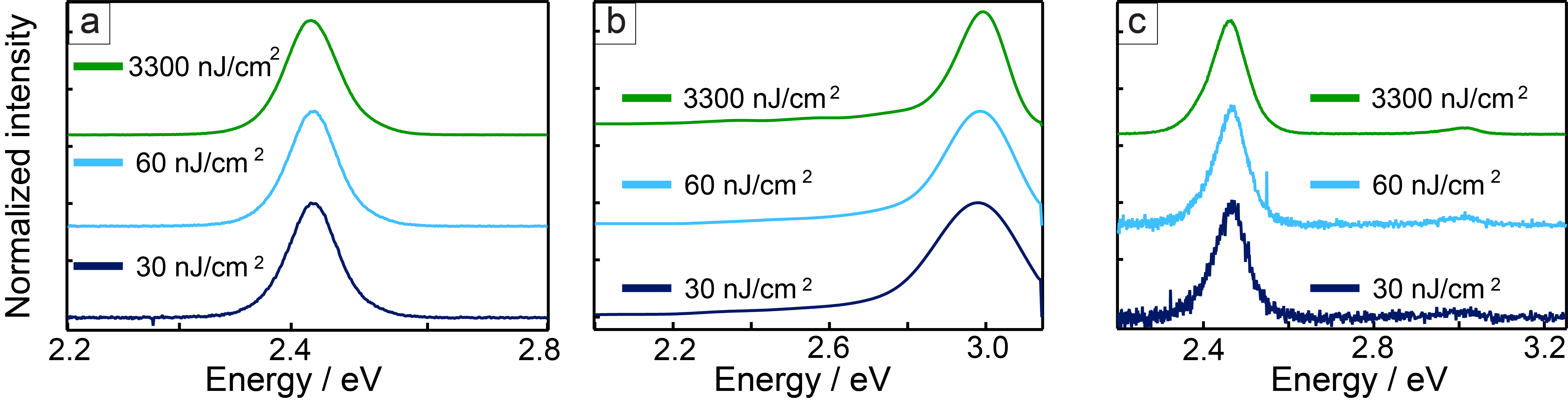}
	\caption{Fluence dependent photoluminescence spectra of (a) one pristine superlattice, (b) one pristine 2DLP microcrystal and (c) a heterostructure.}
	\label{SI_figure:fig_abs}
\end{figure*}
\FloatBarrier
\section{Calculation of biexciton lifetimes}

\begin{figure*}[!htbp]
	\centering
	\includegraphics[width=0.95\textwidth]{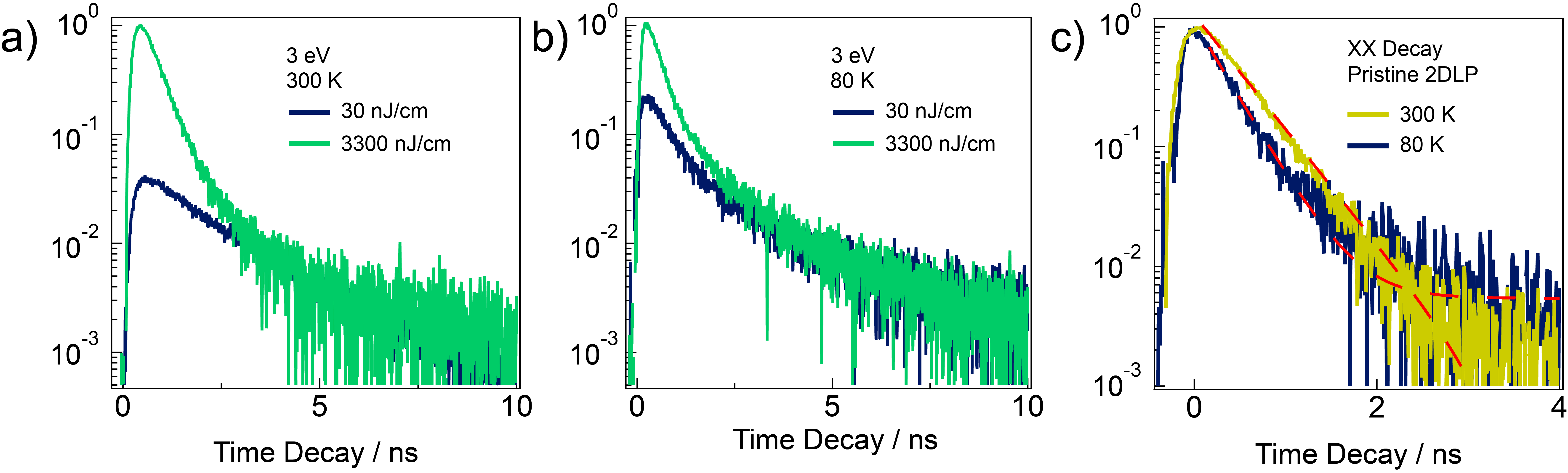}
	\caption{a,b) Time resolved photoluminescence decays measured for pristine 2DLP microcrystals in the low and high fluence regimes at 300 K and 80 K respectively. c) Biexciton decays that were extracted by calculating the difference between the low and high fluence traces in a and b.}
	\label{SI_figure:fig_BiEx_RP}
\end{figure*}

The 2DLP microcrystal biexciton lifetimes $\tau_{2DLP_{XX}}$ were calculated by normalizing the 2DLP microcrystal time-resolved emission measured in the low and high fluence regimes (60  nJ/cm\textsuperscript{2} and 3300 nJ/cm\textsuperscript{2} respectively) at long decay times, where again the emission is supposed to be single excitonic. Following this procedure, we calculated $\tau_{2DLP_{XX}} =$ 430 ps and 350 ps at 293 K and 80 K respectively (\textbf{Figure \ref{SI_figure:fig_BiEx_RP}}). As observed at room temperature, at 80 K in the high fluence regime (3300 nJ/cm\textsuperscript{2}) the biexciton contribution is mitigated in the heterostructure case, as a part of the excitation is channeled to the superlattice domain (see Figure \ref{SI_figure:fig_rp_80k_DECAY}).\\

\FloatBarrier

\begin{figure*}[!htbp]
	\centering
	\includegraphics[width=0.95\textwidth]{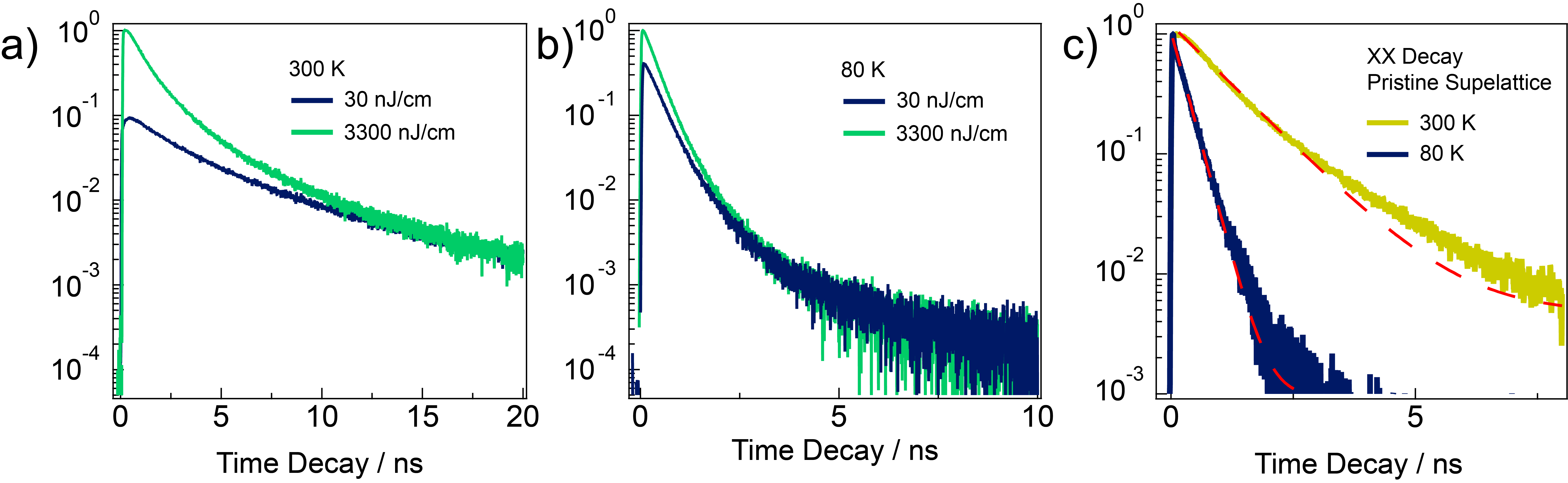}
	\caption{a,b) Time resolved photoluminescence decays measured for pristine superlattices in the low and high fluence regimes respectively. c) Biexciton decays that were extracted by calculating the difference between the low and high fluence traces reported in a and b.}
	\label{SI_figure:fig_BiEx_SL}
\end{figure*}

\FloatBarrier
\pagebreak
\section{Time-resolved emission decay fitting}

\begin{figure*}[!htbp]
	\centering
	\includegraphics[width=0.95\textwidth]{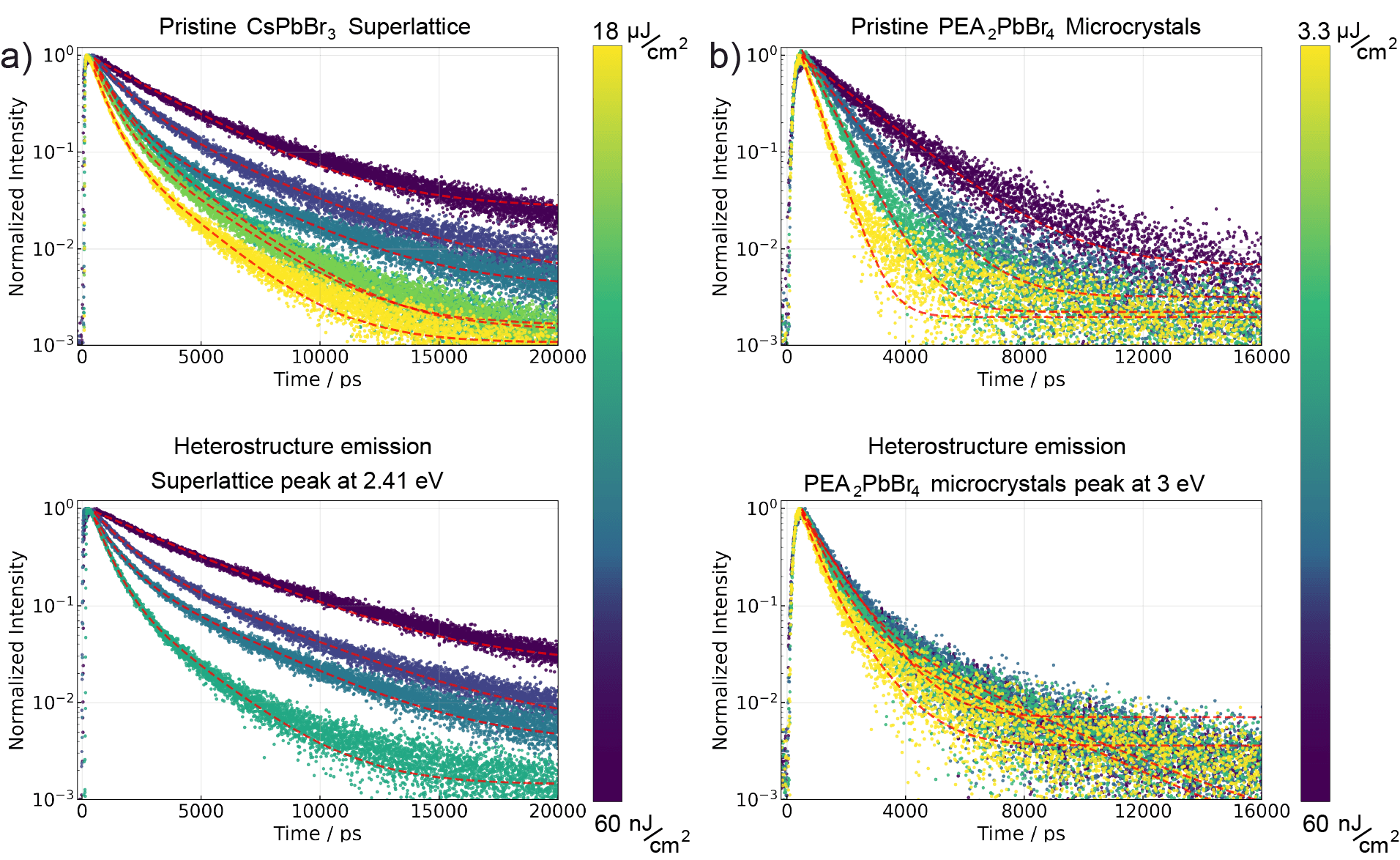}
	\caption{a,b) Time resolved photoluminescence decays measured in correspondence of the CsPbBr$_3$ nanocrystal superlattice emission (a) and of the PEA$_2$PbBr$_4$ 2DLP microcrystal emission (b) in the case of pristine crystals (top) and of heterostructures (bottom). The decay traces measured at the lowest fluence (60 nJ cm$^{-2}$, darkest violet traces) are fitted using a single exponential decay function, all the other traces are fitted by mean of equation \ref{eq:biexp_2}.}
	\label{SI_figure:fig_biexp_fit}
\end{figure*}

\FloatBarrier

The high fluence time-resolved emission decays were fitted using a bi-exponential function expressed as:
\begin{equation}
	I(t) = A_1 \exp{\frac{-t}{\tau_1}} + A_2 \exp{\frac{-t}{\tau_2}} + c,
	\label{eq:biexp_2}
\end{equation}
where $A_n$ is the amplitudes of the exponential decays and $\tau_n$ are the corresponding decay time constants. The fits are shown in Figure \ref{SI_figure:fig_biexp_fit}. The the effective time decay $\tau_{eff}$, reported in Figure 3 of the main text, were calculated using the following expression:
\begin{equation}
	\tau_{eff} = \tau_1 \times (\frac{A_1}{A_1 + A_2}) + \tau_2 \times (\frac{A_2}{A_1 + A_2}).
	\label{eq:biexp}
\end{equation}

\section{Cryogenic optical characterization}

\begin{figure*}[!htbp]
	\centering
	\includegraphics[width=0.95\textwidth]{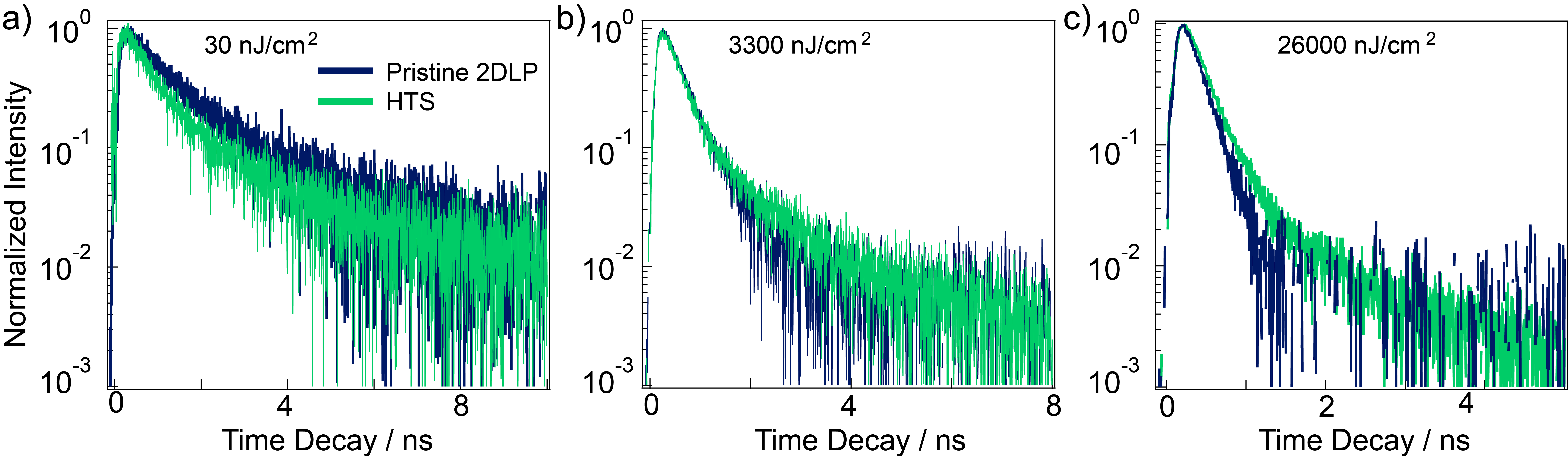}
	\caption{a-c) Time resolved photoluminescence decays measured for pristine 2DLP microcrystals .}
	\label{SI_figure:fig_rp_80k_DECAY}
\end{figure*}

\FloatBarrier

\begin{figure*}[!htbp]
	\centering
	\includegraphics[width=0.95\textwidth]{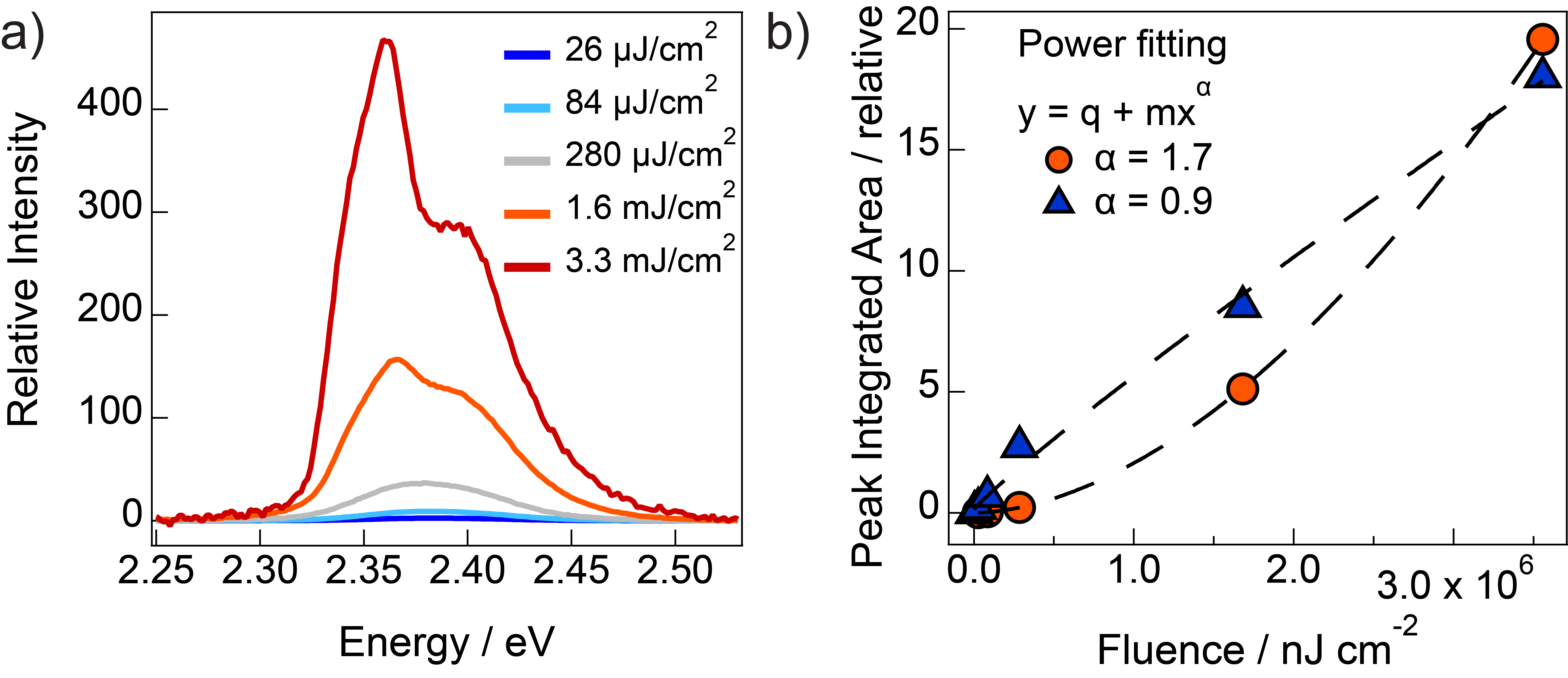}
	\caption{a) Photoluminescence of CsPbBr$_3$ pristine superlattice as a function of fluence. b) Integrated area of high energy (blue triangles) and low energy (red circles) peaks as a function of fluence. Dashed lines represent the power law fitting.}
	\label{SI_figure:s}
\end{figure*}

\FloatBarrier

\begin{figure*}[!htbp]
	\centering
	\includegraphics[width=0.95\textwidth]{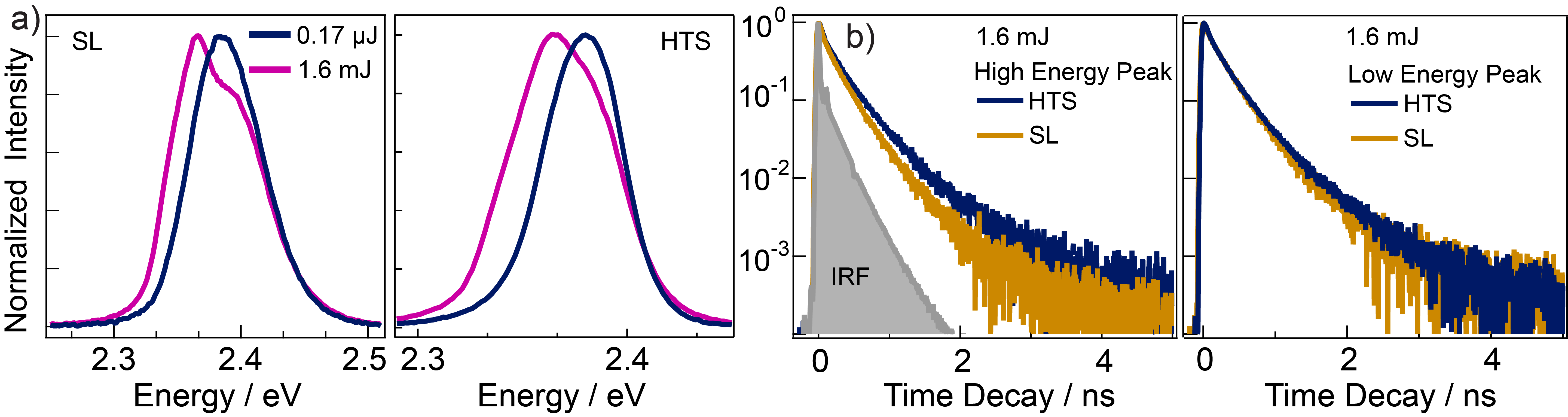}
	\caption{a) Comparison of the emission spectra in the low (170 nJ/cm$^2$) and high (1.6 mJ/cm$^2$) fluence regimes for a pristine superlattice (left graph) and a superlattice assembled in a heterostructure b) Time-resolved emission decays of pristine superlattice and heterostructure extracted for the high (left) and low (right) energy emission peaks.}
	\label{SI_figure:s}
\end{figure*}

\FloatBarrier

\newpage
	
	\bibliographystyle{apsrev4-2}
	\bibliography{A_SI_refs}